\documentclass[11pt]{article}
\usepackage[table]{xcolor}

\usepackage{graphicx}
\usepackage{authblk}
\RequirePackage[margin=1in]{geometry}
\RequirePackage{amsmath,amssymb,amsthm,mathtools}
\RequirePackage{bm}
\RequirePackage{booktabs}
\RequirePackage{tabularx}
\RequirePackage{array}
\RequirePackage{float}
\RequirePackage{enumitem}
\RequirePackage{microtype}
\usepackage[square,authoryear]{natbib}
\RequirePackage{xcolor}
\RequirePackage{url}
\RequirePackage{hyperref}
\RequirePackage[nameinlink,capitalise]{cleveref}
\usepackage{bbm}
\usepackage[ruled,vlined,linesnumbered]{algorithm2e}
\hypersetup{colorlinks=true,linkcolor=black,citecolor=black,urlcolor=black}
\newtheorem{theorem}{Theorem}[section]
\newtheorem{proposition}[theorem]{Proposition}
\newtheorem{lemma}[theorem]{Lemma}
\newtheorem{corollary}[theorem]{Corollary}
\newtheorem{assumption}[theorem]{Assumption}
\newtheorem{definition}[theorem]{Definition}
\newtheorem{remark}[theorem]{Remark}

\newcolumntype{Y}{>{\raggedright\arraybackslash}X}
\newcommand{\R}{\mathbb R}
\newcommand{\N}{\mathbb N}
\newcommand{\X}{\mathcal X}
\newcommand{\Z}{\mathcal Z}

\newcommand{\OO}{\mathcal O}
\newcommand{\A}{\mathcal A}
\newcommand{\Ch}{\widehat{C}_n}
\newcommand{\ch}{\widehat{C}}
\newcommand{\Piv}{\mathcal P_{\mathrm{IV}}}
\newcommand{\Pobs}{\mathcal P_{\mathrm{obs}}}
\newcommand{\One}[1]{{\mathbbm{1}}\left\{{#1}\right\}}
\newcommand{\EE}[1]{\mathbb E\!\left[#1\right]}
\newcommand{\EEst}[2]{\mathbb E\!\left[#1\mid #2\right]}
\newcommand{\Ep}[2]{\mathbb E_{#1}\!\left[#2\right]}

\newcommand{\PP}[1]{\mathbb P\!\left\{#1\right\}}
\newcommand{\PPst}[2]{\mathbb P\!\left\{#1\mid #2\right\}}
\newcommand{\Pp}[2]{\mathbb P_{#1}\!\left\{#2\right\}}
\newcommand{\Ppst}[3]{\mathbb P_{#1}\!\left\{#2\mid #3\right\}}
\newcommand{\leb}[1]{\operatorname{Leb}\!\left(#1\right)}

\DeclareMathOperator*{\argmax}{arg\,max}

\newcommand{\indep}{\mathrel{\perp\!\!\!\perp}}
\newcommand{\notindep}{\mathrel{\not\!\perp\!\!\!\perp}}
\newcommand{\iidsim}{\stackrel{\mathrm{i.i.d.}}{\sim}}
\newcommand{\B}{\mathcal{B}}
\newcommand{\eps}{\varepsilon}

\definecolor{ed}{RGB}{225,100,0}

\begin{document}

\title{Distribution-free predictive inference with instrumental variables}
\date{\today}
\author{Yonghoon Lee}
\author{Edgar Dobriban}
\author{Eric Tchetgen Tchetgen\thanks{E-mail addresses:
\texttt{yhoony31@wharton.upenn.edu},
\texttt{dobriban@wharton.upenn.edu},
\texttt{ett@wharton.upenn.edu}
}}
\affil{Department of Statistics and Data Science,\\
The Wharton School, University of Pennsylvania}

\maketitle

\begin{abstract}
We study the problem of predictive inference with outcomes missing not at random (MNAR), when instrumental variables are available. While conformal prediction and weighted conformal prediction provide valid distribution-free inference when outcomes are missing completely at random and missing at random, respectively, inference is generally not possible without additional assumptions in the MNAR setting, as the distribution of observed and missing outcomes can differ arbitrarily. 
In this work, we leverage instrumental variables (IV), which have been powerful tools in causal inference and missing data.
We show that the possibility of informative predictive inference leveraging the IV is controlled
by a parameter that we refer to as the \textit{identification index}.
Specifically, all distribution-free approaches fail to provide informative inference
when the identification index is below the target coverage level.
In the complementary case, we introduce a practical method that achieves the target coverage
by estimating coverage probabilities and tuning conformal-type prediction sets accordingly.
 While the coverage functional can fail to be pathwise differentiable and thus present potential challenges for estimation, we address this issue by leveraging a policy-based upper bound on the coverage that we then estimate in a doubly robust manner. 
We demonstrate the performance of the proposed procedure through simulations and two empirical datasets.
\end{abstract}



\section{Introduction}
Consider a predictive inference problem with missing outcomes, where one observes data points $(X_i,R_i,R_iY_i)$ for $i=1,\ldots,n$---with the outcome $Y_i$ observed when $R_i=1$ and unobserved when $R_i=0$.
Given a new feature input $X_{n+1}$, the goal is to construct a prediction set $\ch(X_{n+1})$ for the corresponding unobserved outcome $Y_{n+1}$.
In distribution-free inference, we aim to develop a method for this problem without imposing distributional assumptions, such as smoothness or discreteness, that may be difficult to verify, while working under general conditions---such as i.i.d. or exchangeability---to attain reliable inference.

The difficulty of distribution-free inference with missing outcomes depends heavily on the missingness pattern. For example, if the outcomes are \textit{missing completely at random}, i.e.,
the missingness indicator $R$ satisfies
$R \indep (X,Y)$, one can apply conformal prediction~\citep{vovk2005algorithmic} to the fully observed datapoints $(X_i,Y_i)_{i:R_i=1}$ to construct a prediction set with finite-sample coverage.
If the outcomes are \textit{missing at random} (MAR), i.e., $R \indep Y\mid X$, one can instead apply weighted conformal prediction~\citep{tibshirani2019conformal}, which provides a valid prediction set when the propensity score (i.e., the probability of observing the outcomes given the features) is known. 
When the propensity score is unknown, exact distribution-free coverage control becomes difficult, 
but one can attain certain forms of approximate inference with robust coverage guarantees~\citep{lei2021conformal,yang2024doubly,qiu2022prediction}.

When the outcome is \textit{missing not at random} (MNAR), i.e., $R \notindep Y\mid X$, distribution-free inference essentially becomes impossible, since the distribution of the observed datapoints $P_{(X,Y)\mid R=1}$ can be arbitrarily different from that of the test point $P_{X,Y}$.
This prevents us from sufficiently learning about the test distribution from the calibration data.
Therefore, additional information is needed to enable valid inference. For example, a classical approach is to impose parametric assumptions on the outcome and missingness mechanisms, as exemplified by the sample selection model of \citet{heckman1979sample}.

In this work, we explore the MNAR setting where additional information is available in the form of instrumental variables (IVs). A missing data IV is a fully observed variable that is associated with the missingness mechanism but is conditionally independent of the outcome given the covariates in the underlying population of interest. This structure provides information that can help account for the dependence between missingness and the outcome. We investigate how IVs can be leveraged for predictive inference.

For example, consider HIV testing in a population-based survey, where $X$ represents individual characteristics, $R$ denotes whether an individual undergoes HIV testing, and $Y$ denotes the individual's HIV status. The outcome is observed only when $R=1$, i.e., when the individual undergoes HIV testing. Suppose each individual is assigned an interviewer with identity $Z$ who conducts a survey interview and seeks consent for HIV testing. After the interview, the individual decides whether to undergo testing. In this setting, the interviewer's characteristics, such as years of experience as a health survey worker, may affect an interviewee's decision to undergo HIV testing $R$. However, it is unlikely that the individual's HIV status depends on which interviewer is assigned. Thus, interviewer identity can serve as an IV to address missing outcomes, as illustrated in studies of HIV testing in Zambia \citep{mcgovern2015using}. See Section~\ref{sec:literature} for further discussion of the literature on IV methods.

Instrumental variables have been extensively studied in the literature on missing data and causal inference, primarily in the context of estimation and inference for a given population functional of interest; see, e.g., \citet{angrist1996identification,tchetgen2017general}. Motivated by their widespread use and success in these areas, we investigate how IVs can be leveraged for predictive inference.
Specifically, we consider the setting where we observe data points $(X_i,Z_i,R_i,R_iY_i)_{i=1,\ldots,n}$ and aim to perform predictive inference on $Y_{n+1}$ for a new observation $X_{n+1}$, under the IV assumption $Z \indep Y \mid X$.

The first question is to understand when nontrivial predictive inference (e.g., finite prediction sets for continuous real-valued outcomes) is possible. In general, having an IV is not sufficient to identify different target parameters of interest, and additional assumptions are typically required, e.g.,~\citep{angrist1995identification,wang2018bounded,liu2025multiplicativeinstrumentalvariablemodel,hernan2006instruments}. In particular, classical partial identification bounds on the mean are necessarily trivial for outcomes with unrestricted support when part of the outcome distribution remains unidentified~\citep{manski1989anatomy,manski2003partial}. 
We first show that analogous fundamental limitations also arise in predictive inference: when a new parameter that we call the \textit{identification index} is below the target coverage level, any valid distribution-free prediction set must have infinite width on average and is therefore uninformative.

Intriguingly, however, there is also a regime where nontrivial predictive inference is possible without imposing additional assumptions beyond IV validity, even for unbounded outcomes: this occurs when the identification index is above the target coverage level. In this regime, we develop a methodology for constructing prediction sets with approximate coverage guarantees, where the bound on potential undercoverage is doubly robust to estimation errors in nuisance functions. The identification index has a transparent interpretation: it measures how much of the outcome distribution can be captured by combining the information available across different values of the instrument.

\paragraph{Notation}
For a positive integer $n$, we write $[n]$ to denote the set $\{1,2,\ldots,n\}$. We write $\mathcal B(\R)$ to denote the Borel $\sigma$-algebra on the real space $\R$. For a measurable space $\X$, we write $\mathcal{P}(\X)$ to denote the set of all probability measures on $\X$. For a measurable set $A\subset\R$, we write $\leb{A}$ to denote its Lebesgue measure. For a set $A$ and a positive integer $n$, $A^n$ denotes the set of vectors of length $n$ with elements in $A$. For two sets $A$ and $B$, $B^A$ denotes the set of functions from $A$ to $B$, i.e., $B^A = \{f:A\rightarrow B\}$.

\subsection{Main contributions}
We develop theory and methodology for distribution-free predictive inference in the IV setting. Our main contributions can be summarized as follows.

\paragraph{Complete characterization of the possibility of distribution-free inference with an instrument}
We establish that the possibility of meaningful distribution-free inference in the IV setting is characterized by a parameter that we call the \textit{identification index}. We prove an impossibility result in the regime of an identification index below a threshold set by the desired coverage level, while for an identification index above the threshold, we show that meaningful IV based predictive inference is theoretically possible without additional assumptions.  Notably, meaningful predictive inference remains possible even with unbounded outcomes, for which classical partial identification bounds on the mean are necessarily trivial unless the outcome distribution is fully identified~\citep{manski1989anatomy}.

\paragraph{Doubly robust predictive inference with IV}

For the regime where inference is possible, we develop a practical procedure based on semiparametric theory. We first introduce an oracle procedure that achieves exact distribution-free coverage in the infinite-sample regime. However, its practical implementation faces two challenges: 
first, a plug-in procedure suffers from first-order nuisance estimation bias, potentially causing substantial over- or under-coverage. 
Second, the oracle functional may fail to be pathwise differentiable due to pointwise maximization, making influence-function-based bias correction difficult. To address these challenges, we introduce a policy-based approach (analogous to the ones from \cite{vanderlaan2015targeted} in a different setting of dynamical treatment rules) that admits an efficient influence function with a second-order, doubly robust coverage bias (Section~\ref{sec:method_policy}). We establish theoretical coverage guarantees and demonstrate that the resulting procedure achieves well-calibrated coverage when the identification index is sufficiently large.

\paragraph{Empirical evaluation}
We illustrate the performance of the proposed procedure through simulations and empirical experiments using two datasets, the 
FES wage data and the NHANES III self-reported height data.\footnote{Code to reproduce the experiments is available at \url{https://github.com/yhoon31/distribution_free_IV}.} 
The results illustrate that our method generally achieves coverage around or above the desired level and produces informative prediction intervals across a range of settings.

\subsection{Related work}\label{sec:literature}

Distribution-free inference and conformal prediction have gained much interest in recent literature. For overviews of results in this field, see~\citet{vovk2005algorithmic,angelopoulos2021gentle} and~\citet{angelopoulos2024theoretical}. Conformal prediction (see, e.g., \citet{saunders1999transduction,vovk1999machine,papadopoulos2002inductive}) provides a distribution-free framework for generating valid prediction sets under exchangeability.

Many works have tried to extend conformal prediction beyond exchangeable data. For the covariate shift setting, \citet{tibshirani2019conformal} introduced weighted conformal prediction, which was further studied in~\citet{lei2021conformal}. \citet{yang2024doubly} provided a method with a doubly robust coverage gap. See also~\citet{qiu2022prediction} and~\citet{lee2025conditionalpredictiveinferencemissing}. Related work on prediction sets under dataset shift includes PAC prediction sets under covariate shift~\citep{park2021pac} and label shift~\citep{si2023pac}, as well as conformal inference under high-dimensional covariate shifts via likelihood-ratio regularization~\citep{joshi2025conformal}. \citet{doi:10.1073/pnas.2214889120} studied the setting where the distribution shift occurs not only in the covariates but also in the label. They propose a method that provides valid coverage under a hypothetical constraint on the influence of unmeasured confounding. \citet{yi2025conformal} develop weighted conformal prediction methods for MNAR outcomes by estimating the missingness propensity under additional modeling assumptions.

Instrumental variable (IV) methods have a long history in econometrics, dating back to the seminal work of \citet{wright1928tariff} on simultaneous equation models; see also \citet{goldberger1972structural}. Several studies have investigated the use of instrumental variables (IVs) to address outcomes missing not at random. Adjusting for non-random selection or missing data using exclusion restrictions is a well-established practice across the social sciences, especially in econometrics~\citep{heckman1979sample,dubin1989selection,winship1992models,manski2003partial,nicoletti2010poverty, das2003nonparametric}.
 In the health sciences, \citet{barnighausen2011correcting} employ a Heckman-type parametric selection model to correct for nonresponse bias in estimating HIV prevalence. \citet{geneletti2011adjusting} examined the sensitivity of this approach to modeling assumptions, while \citet{clark2014validation} further evaluated its robustness. \citet{mcgovern2015assumption} proposed a copula-based selection model to relax the bivariate normality assumption.

Beyond parametric selection models, \citet{tchetgen2017general} established nonparametric identification of population regression functions under an additional restriction on selection bias, while \citet{sun2018semiparametric} established identification conditions for the full data distribution under MNAR using an IV and developed semiparametric estimators for the population mean. For more general references on semiparametric theory and estimation, see \citet{levit1974optimality,Bickel1993,tsiatis2006}. In a different direction, \citet{marden2018implementation} developed practical methods for computing IV-based bounds on population proportions and testing for selection bias under MNAR. \citet{kato2026conformaliv} develops conformal prediction methods for the nonparametric instrumental variable model to address endogeneity between predictors and outcome variables, under a separable structural equation model---rather than missing outcomes. 
See also~\citet{newey2003instrumental,ai2003efficient,darolles2011nonparametric} for related work on nonparametric instrumental variable estimation.

As previously mentioned, instrumental variables have also been extensively studied in causal inference, particularly for inference on causal effects under unmeasured confounding---see, e.g.,~\citet{angrist1996identification,levis2024nonparametric,swanson2018partial}. 
Our focus, however, is on predictive inference for missing outcomes rather than causal effects.

To our knowledge, these two strands of literature, i.e. conformal prediction and instrumental variables, have largely developed independently, and we are not aware of any existing methods for distribution-free predictive inference that leverage instrumental variables to address missing outcomes. Our work aims to bridge this gap.

\section{Problem setting}

Suppose we have observations
$(X_i, Z_i, R_i, Y_i R_i)_{i \in [n]}$, where the underlying data $(X_i, Z_i, R_i, Y_i)_{i \in [n]}$ are drawn i.i.d.\ from an unknown distribution $P = P_X \times P_{Z \mid X} \times P_{R \mid Z, X} \times P_{Y \mid X,Z,R}$ on $\X \times \Z \times \{0,1\} \times \R$.
The outcome $Y_i$ is observed only when $R_i = 1$ and not when $R_i=0$, which we encode in the observation being $Y_iR_i$. 
Here, $Z_i$ denotes the potential instrumental variable for the $i$th unit.
We consider the setting where the instrumental variable $Z$ is categorical, i.e., $|\Z| < \infty$.

\begin{assumption}[Instrumental variable]\label{asm:iv}
    The random variables $(X,Z,Y)$ satisfy $Z \indep Y \mid X$ under the distribution $P$.
\end{assumption}
Assumption~\ref{asm:iv} specifies the IV condition imposed on the data distribution and is the only IV assumption required throughout this paper. In particular, we do not require the conventional IV relevance condition $Z \not\indep R \mid X$ to develop our theory. Nevertheless, our framework accommodates settings where this condition holds, and all subsequent theoretical results remain valid without modification if it is additionally imposed.

Now, given a new feature input $X_{n+1} \sim P_X$, we aim to construct a prediction set $\ch(X_{n+1})$ for the corresponding outcome $Y_{n+1}$, with the coverage guarantee
\[\PP{Y_{n+1} \in \ch(X_{n+1})} \geq 1-\alpha,\]
where $\alpha \in (0,1)$ is a prespecified target level, and the probability is taken with respect to $(X_i, Z_i, R_i, Y_i)_{i \in [n+1]} \iidsim P$.\footnote{
While we write $(X_i,Z_i,R_i,Y_i)_{i\in[n+1]}\iidsim P$ for simplicity, only $X_{n+1}$ and $Y_{n+1}$ need to be 
defined for the test point.}

 In classical predictive inference and conformal prediction, such a coverage guarantee is the most commonly studied goal \citep{saunders1999transduction,papadopoulos2002inductive,vovk2005algorithmic}. 
In the presence of missing data, 
this guarantee and its approximate versions 
are also known to be achievable under missing at random assumptions~\citep{tibshirani2019conformal,yang2024doubly}.

\subsection{Distribution-free predictive inference with instrumental variables}

Here, we formalize the meaning of distribution-free inference in the setting with instrumental variables. Let
\[\Piv=\Big\{\,P\in\mathcal P(\X\times\Z\times\{0,1\}\times\R): Z\indep Y\mid X\text{ under }P\, \Big\}\]
denote the set of all joint distributions of $(X,Z,R,Y)$ satisfying  Assumption~\ref{asm:iv}, and let $\OO = \X\times\Z\times\{0,1\}\times \R$ be the space of observations. The notion of distribution-free prediction set can be described as follows.

\begin{definition}\label{def:df_cov}
An algorithm
\(
\A:\OO^n\to\B(\R)^\X
\)
is said to satisfy \emph{distribution-free coverage under IV} at level \(1-\alpha\) if, for every distribution $P \in \Piv$,
\[
\Pp{(X_i,Z_i,R_i,Y_i)_{i\in[n+1]}\iidsim P}
{Y_{n+1}\in\Ch(X_{n+1})}
\ge 1-\alpha,
\]
where
\(
\Ch=\A\bigl((X_i,Z_i,R_i,R_iY_i)_{i\in[n]}\bigr)
\)
denotes the prediction set function obtained by applying algorithm $\A$ to the observed data.
\end{definition}

Although Definition~\ref{def:df_cov} gives a seemingly straightforward notion of distribution-free coverage, there are several additional considerations that arise when seeking useful predictive inference under this requirement. In particular, while the coverage guarantee should hold uniformly over the entire distribution class, the prediction set itself should still be adaptive to the underlying distribution and become more informative when the data distribution enables that. 
For example, a randomized procedure that outputs $\R$ with probability $1-\alpha$ and the empty set with probability $\alpha$ satisfies the desired coverage guarantee for every distribution without using the data at all, but is not considered useful. In this regard, our goal is to construct a procedure that provides distribution-free validity while also retaining distribution-adaptive informativeness, in a way that we make precise below. 
These two goals are inherently in tension, making the problem substantially more challenging.

In the missing-outcome setting, there are two distinct challenges to achieving adaptivity to the distribution. 
First, even if the distribution of the observed data were known, 
the missing outcome distribution would only be partially identified.
Since our goal is predictive inference for these missing outcomes, 
the amount we could learn from the data could still vary substantially across underlying distributions.
Also, in practice, the distribution of the observed data can only be characterized using an available finite sample, which can result in additional estimation error that can harm coverage.

To disentangle these two sources of difficulty, we first consider an \emph{oracle} setting in which the procedure has access to the distribution of the observed data. This removes the effect of estimation error and allows us to first understand the extent to which a distribution-free prediction set can adapt to the distribution of the observed data. This setting can roughly be viewed as addressing the infinite-sample regime. We then turn to the practical (non-oracle) setting, where the relevant nuisance functions must be estimated from finite data.

A similar distinction between oracle and estimated quantities arises in prior distribution-free methods for missing outcomes in the MAR setting. 
For example, weighted conformal prediction~\citep{tibshirani2019conformal} achieves exact finite-sample coverage when the relevant likelihood ratio or propensity score is known, while estimating these quantities generally introduces additional coverage error \citep{lei2021conformal,yang2024doubly}. 

To distinguish between oracle and non-oracle procedures,
we now define the distribution-free coverage property for an oracle procedure that leverages the distribution of the observed data. Let
\begin{equation}\label{pobs}
    \Pobs=\left\{P'\in\mathcal P(\OO): P'=P_{X,Z,R,RY}\text{ for some }P\in\Piv\right\}
\end{equation}
be the set of distributions of the observed data under the IV model.

\begin{definition}\label{def:df_cov_oracle}
An \emph{oracle algorithm}
\(
\A:\OO^n \times \mathcal{P}_{\rm obs}\to\B(\R)^\X
\)
is said to satisfy \emph{distribution-free coverage under the IV conditions} at level \(1-\alpha\) if, for every distribution $P \in \Piv$,
\[
\Pp{(X_i,Z_i,R_i,Y_i)_{i\in[n+1]}\iidsim P}
{Y_{n+1}\in\Ch(X_{n+1})}
\ge 1-\alpha,
\]
where
\(
\Ch=\A\bigl((X_i,Z_i,R_i,R_iY_i)_{i\in[n]}, P_{X,Z,R,RY}\bigr),
\)
and \(P_{X,Z,R,RY}\) denotes the distribution of \((X,Z,R,RY)\) induced by \(P\).
\end{definition}
We allow algorithms to use an auxiliary random seed independent of the data, which is understood to be included in all probabilities and expectations. Note that an algorithm that has access to only the empirical data can be viewed as a special type of oracle algorithm that does not use oracle information. 
 
We first characterize when an oracle procedure with distribution-free coverage is possible (Section~\ref{sec:theory}) and then construct a 
procedure in the feasible regime (Section~\ref{sec:methodology}).
 We present 
 experimental results in Sections~\ref{sec:simulation} and~\ref{sec:empirical-data}.

\section{Possibility of distribution-free inference with IV}\label{sec:theory}

\subsection{Overview}

In this section, we provide a theoretical analysis of when informative distribution-free predictive inference is possible in the IV setting---where ``informative'' roughly means that the prediction set has a finite width. 

We show that, for each IV distribution $P \in \Piv$, there exists a parameter $\Gamma_P$---which we refer to as the \emph{identification index}--- that characterizes the possibility of informative inference. 
We establish an impossibility result showing that, when $\Gamma_P$ is below the target coverage level, no distribution-free inference procedure can provide informative inference. Conversely, when $\Gamma_P$ exceeds this threshold, informative inference is possible. We develop these results in the following subsections.

\subsection{Identification index}

To motivate the identification index, fix an IV distribution $P\in\Piv$ and a covariate value $x$. For ease of exposition, we first consider the continuous outcome case, where the conditional outcome distributions admit densities. By the IV assumption $Y\indep Z\mid X$, all instrument groups share the same underlying outcome density $f_{Y\mid X=x}$. 
However, missingness restricts 
us to only observe part of this density in each group: at an outcome value $y$, the fraction retained in group $z$ is $\Pp{P}{R=1\mid Y=y,Z=z,X=x}$. 
Taking \emph{the most favorable instrument} at each outcome value therefore gives the retained portion
\begin{equation*}
    i_{P,x}(y)=f_{Y\mid X=x}(y)\cdot\max_{z\in\Z}\Pp{P}{R=1\mid Y=y,Z=z,X=x}.
\end{equation*}
Importantly, the maximizing instrument may depend on $y$, so different instrument groups can reveal different parts of the outcome distribution.  

For example, suppose that there is a perfect instrument value $z_\star$ at which everyone with covariates $x$ responds: $\Pp{P}{R=1\mid Z=z_\star,X=x}=1$. In this case, the respondent outcome density coincides with $f_{Y\mid X=x}$, and hence no probability mass is lost. More generally, even without a single perfect instrument, different instrument values may collectively reveal the entire outcome distribution by ensuring response in different regions of the outcome space. This motivates measuring the total retained mass after combining information across instrument groups.

We now give an equivalent definition in terms of quantities
identifiable from the observed data distribution. 
For each $z\in\Z$ and $x\in\X$, 
define
$\pi_{z,x}=\Ppst{P}{R=1}{Z=z,X=x}$
to be the response rate in instrument group $z$.
Also,
let
\(
F_{z,x,1}=P_{Y\mid R=1,Z=z,X=x}
\)
and $f_{z,x,1}$
be the distribution and density of $Y\mid R=1,Z=z,X=x$,  the outcome among respondents in that group, 
respectively,
whenever $\pi_{z,x}>0$. By Bayes' rule and $Y\indep Z\mid X$, we have
\begin{equation}
    \label{n1}
\pi_{z,x}f_{z,x,1}(y)=f_{Y\mid X=x}(y)\cdot\Pp{P}{R=1\mid Y=y,Z=z,X=x}.
\end{equation}
Thus, $i_{P,x}$ can equivalently be written directly in terms of the observed data, as
$i_{P,x}(y)=\max_{z\in\Z}\pi_{z,x}f_{z,x,1}(y)$.

In order to measure how much of the distribution of the outcome is identified for a specific $x$, we integrate this quantity over the real line and define 
\(\gamma_P(x)=\int_{\R}i_{P,x}(y)\,\mathrm{d}y.
\)
Based on our formulation above this is also equivalent to 
\begin{equation}\label{gammap}
    \gamma_P(x)=\Ep{Y\sim P_{Y\mid X=x}}{\max_{z\in\Z}\Pp{P}{R=1\mid Y,Z=z,X=x}}.
\end{equation}
Thus, $\gamma_P(x) \in[0,1]$ 
represents the average probability of observing the outcome under the most favorable instrument value for each outcome. In particular, $\gamma_P(x)=1$ if and only if, for $P_{Y\mid X=x}$-almost every $y$, at least one instrument value ensures response with probability one. In summary, $\gamma_P(x)$ can be viewed as measuring the probability mass of the underlying outcome distribution captured by combining information across instrument groups.

We then define the \emph{identification index} of $P$ as the expected value of this quantity over the feature distribution.
\begin{definition}[Identification index]
    For a distribution of observed data as described in equation \eqref{pobs},
    recalling $\gamma_P$ from \eqref{gammap}, 
    the identification index is defined as 
\begin{equation}
\Gamma_P=\Ep{X\sim P_X}{\gamma_P(X)}.
\end{equation}
\end{definition}

The identification index turns out to be a fundamental notion that controls the possibility and impossibility of predictive inference in the IV setting.\footnote{ 
The above definition extend naturally to arbitrary outcome spaces by replacing the Lebesgue densities with Radon--Nikodym derivatives with respect to a common dominating measure---see Appendix~\ref{sec:index_general}. Since this extension is purely measure-theoretic and does not affect the statistical arguments, we focus on the continuous case throughout the paper for simplicity.}

We note that the identification index is closely related to the literature on partial identification of outcome distributions under IV assumptions. In particular,~\citet{manski2003partial} and~\citet{kitagawa2021identification} characterize the sharp identified region through the pointwise maximum of instrument-specific observed subdensities in settings without covariates. The identification index extends this construction by conditioning on covariates and allowing instruments with arbitrary finite support, which allows us to study the predictive inference problem---where the goal is to construct prediction sets for the outcome $Y$ given covariates $X$.

\subsection{An impossibility result}

We show that if the identification index $\Gamma_P$ is smaller than $1-\alpha$, then \emph{no algorithm with 
distribution-free coverage guarantee---not even an oracle algorithm with access to the true observation distribution---can attain informative inference.} This underlies the fundamental nature of our identification index. 

\begin{theorem}\label{thm:impos}
Suppose an oracle algorithm $\A:\OO^n \times \mathcal{P}_{\rm obs}\to \B(\R)^\X$ satisfies distribution-free coverage under IV at level $1-\alpha$ in the sense of Definition~\ref{def:df_cov_oracle}. Then, for any distribution $P \in \Piv$ satisfying $\Gamma_P < 1-\alpha$, it holds that
\begin{equation}\label{i}
\EE{\leb{\Ch(X_{n+1})}} = \infty, \qquad \text{ where } \Ch=\A\bigl((X_i,Z_i,R_i,R_iY_i)_{i\in[n]}, P_{X,Z,R,RY}\bigr).
\end{equation}
The expectation is taken with respect to $(X_i,Z_i,R_i,Y_i)_{i\in[n+1]}\iidsim P$.
\end{theorem}

Theorem~\ref{thm:impos} 
states that 
methods 
attaining distribution-free coverage will not be informative when $\Gamma_P < 1-\alpha$.\footnote{This type of impossibility statement is standard in the distribution-free inference literature; see, e.g.,~\citet{foygel2021limits,lee2021distribution,medarametla2021distribution}; however, to the best of our knowledge, this is the first such result in the IV literature.} It immediately implies the same impossibility result for non-oracle procedures satisfying Definition~\ref{def:df_cov}.

The key idea behind the proof of Theorem~\ref{thm:impos} is to construct, for any $P \in \Piv$, another distribution $P' \in \Piv$ that is indistinguishable from $P$ based on the observed data---i.e., $P_{X,Z,R,RY} = P_{X,Z,R,RY}'$---while being sufficiently different from $P$. The possibility of this construction forces the prediction set to have infinite expected width.

Given this result, we next investigate the possibility of informative inference when $\Gamma_P > 1-\alpha$. We separately discuss the corner case $\Gamma_P = 1-\alpha$ in Appendix~\ref{sec:corner_case}---where we show that both finite and infinite expected widths are possible---although this exact equality is highly unlikely to hold in practice.

\subsection{The regime where inference is possible}

We now show that informative inference is possible when $\Gamma_P > 1-\alpha$, 
by constructing a finite prediction set.
 We first construct an oracle prediction set, which illustrates the phenomenon clearly, and then use this as a guide to construct a practical method. 
Consider an arbitrary fixed non-conformity score $s: \mathcal{X} \times \R \to \mathbb{R}$,
which is usually interpreted as a measure of prediction error.
 For instance, given a pre-trained predictor $\hat \mu : \mathcal{X} \to \mathbb{R}$ that is constructed using separate training data, a standard choice is the absolute residual $s(x,y) = |y-\hat{\mu}(x)|$.
For $t\in\R$, define 
the function $\delta$, taking values
\begin{equation}\label{eqn:delta_def}
\delta(t,x) =  1 - \int_\R i_{P,x}(y) \cdot \One{s(x,y) \leq t} \mathrm{d}y
\end{equation}
for all $t,x$.
 This can be interpreted as an upper bound on the miscoverage rate of a prediction set of the form $x\mapsto \{y: s(x,y) \leq t\}$ conditional on the features $x$.
Due to the definition of $\gamma_P(x)$, and by bounding $\One{s(x,y) \leq t} \le 1$,
 this quantity is in the interval $[1-\gamma_P(x),1]$.
Note also that $\delta(t,x)$ is fully determined
by the distribution $P_{X,Z,R,RY}$ of the observed data. 

We then construct an oracle prediction set (referred to as ``oracle" because it depends on the full distribution of the observed data instead of a finite dataset) 
defined as
\begin{equation}\label{eqn:pred_set_oracle_2}
    \Ch^{\mathrm{or}}(x)=\{y\in\R:s(x,y)\le t_\alpha\}, \text{ where } t_\alpha
= \inf\left\{t\in\R: \Ep{X\sim P_X}{\delta(t,X)}\leq\alpha\right\}.
\end{equation}
 The threshold $t_\alpha$ depends on the function $\delta$ as well as the marginal distribution of the features $X$.

For the estimation of the marginal distribution of the features $X$, it turns out that due to our distribution-free assumptions, it suffices to use the empirical distribution.
 Therefore we also define a semi-oracle procedure, which still uses the oracle miscoverage function $\delta$ but uses the empirical distribution of the features $X$.
 This is defined as
\begin{equation}\label{eqn:pred_set_oracle}
    \Ch^{\mathrm{sor}}(x)=\{y\in\R:s(x,y)\le \hat{t}_\alpha\}, \text{ where } \hat{t}_\alpha=\inf\left\{t\in\R:\frac{1}{n+1}\left(\sum_{i=1}^n\delta(t,X_i)+1\right)\le\alpha\right\}.
\end{equation}

The thresholds $t_\alpha$ and $\hat{t}_\alpha$ are set to $+\infty$ if the corresponding sets are empty. We now show that these methods achieve a valid coverage control.

\begin{theorem}\label{thm:coverage}
    For any distribution $P \in \Piv$, the oracle prediction set $\Ch^{\mathrm{or}}(X_{n+1})$ from~\eqref{eqn:pred_set_oracle_2} and the semi-oracle prediction set $\Ch^{\mathrm{sor}}(X_{n+1})$ from~\eqref{eqn:pred_set_oracle} satisfy
    \[\PP{Y_{n+1} \in \Ch^{\mathrm{or}}(X_{n+1})} \geq 1-\alpha
    \quad \text{ and }  \quad \PP{Y_{n+1} \in \Ch^{\mathrm{sor}}(X_{n+1})} \geq 1-\alpha,\]
    where the probability is taken with respect to $(X_{n+1},Y_{n+1}) \sim P_{X,Y}$ for $\Ch^{\mathrm{or}}$ and $(X_1,Z_1,R_1,Y_1),\ldots,$ $(X_{n+1},Z_{n+1},R_{n+1},Y_{n+1}) \iidsim P$ for $\Ch^{\mathrm{sor}}$.
\end{theorem}

Next, we show that the oracle prediction set $\Ch^{\mathrm{or}}$ in~\eqref{eqn:pred_set_oracle_2} can indeed have finite measure under the condition $\Gamma_P > 1-\alpha$.
To establish that finite average measure is possible, one choice of a score $s$ is enough.
Below, we choose the score function to be the absolute value of the outcome itself, i.e., $s(x,y)=|y|$; while in practice we can of course use other scores.

\begin{theorem}\label{thm:finite_measure}
Suppose that $P\in\Piv$ satisfies $\Gamma_P>1-\alpha$.
Then the oracle prediction set $\Ch^{\mathrm{or}}$ defined
in~\eqref{eqn:pred_set_oracle_2}, with score $s(x,y)=|y|$, satisfies
\begin{equation*}
    \EE{\leb{\Ch^{\mathrm{or}}(X_{n+1})}}<\infty.
\end{equation*}
\end{theorem}

\subsection{A complete characterization}

Combining the results of Theorems~\ref{thm:impos},~\ref{thm:coverage}, and~\ref{thm:finite_measure}, together with Theorem~\ref{thm:corner_case} for the corner case, we obtain the following summary. Let
\[\Omega_\alpha = \{\A:\OO^n \times \mathcal{P}_{\rm obs}\to\B(\R)^\X, \text{ $\A$ satisfies the distribution-free coverage under IV at level $1-\alpha$}\}\]
denote the set of all oracle algorithms that satisfies Definition~\ref{def:df_cov_oracle} at level $1-\alpha$.

\begin{corollary}\label{cor:characterization}
For an IV distribution $P \in \Piv$, the possibility of informative (oracle) distribution-free predictive inference is characterized by the identification index as follows:
\[
\inf_{\A \in \Omega_\alpha} \Ep{P}{\leb{\Ch(X_{n+1})}}
=
\begin{cases}
\infty, & \text{if }\, \Gamma_P < 1-\alpha,\\[0.5ex]
\Lambda_P& \text{if }\, \Gamma_P = 1-\alpha,\\[0.5ex]
<\infty, & \text{if }\, \Gamma_P > 1-\alpha,
\end{cases}
\]
where $\Lambda_P=\Ep{X \sim P_X}{\leb{\{y\in\mathbb R:i_{P,X}(y)>0\}}}$.
\end{corollary}

Clearly, non-oracle prediction sets, which do not have access to the full observed data distribution, can only be as good as or worse than oracle prediction sets. 
Hence, to summarize, these results imply that
there are two general strategies for predictive inference with an IV:
\begin{enumerate}
\item \textit{Completely distribution-free methodology.} 
We may aim to develop a prediction procedure that (approximately) satisfies coverage as per Definition~\ref{def:df_cov}, regardless of the underlying distribution. According to Corollary~\ref{cor:characterization}, however, 
this can be informative only when $\Gamma_P>1-\alpha$.

\item \textit{Imposing additional assumptions.}
When there is large missingness even for the best instrument, i.e., $\Gamma_P<1-\alpha$, 
 then our theory strongly suggests 
that distribution-free inference is not possible
 and we will need to impose additional assumptions.

\end{enumerate}

As discussed above, both strategies have their own advantages and limitations. In this work, we focus on the first strategy, namely completely distribution-free inference. 

\begin{remark}
The possibility of (oracle) inference is separate from how well the instruments are assigned---e.g., whether each patient with covariates $X$ is assigned an ``efficient'' interviewer $Z$ who is likely to yield a high response probability $\PPst{R=1}{Z,X}$.

 Indeed, the identification index does not depend on the distribution of $Z$ given $X$. Instead, it depends only on $i_{P,x}(y)=\max_{z\in\Z}\pi_{z,x}f_{z,x,1}(y)$, which captures the most favorable instrument value at each $(x,y)$. Thus, instrument assignment strategy cannot theoretically improve the possibility of inference. 
 However, for the practical methodology introduced in the next section, the distribution of $Z$ also matters, as it affects our ability to obtain reliable nuisance estimators.
\end{remark}

\begin{remark}
    This characterization implies that informative predictive inference is also possible without an IV if the observation probability is sufficiently large---specifically, when $\PP{R=1}>1-\alpha$. Nevertheless, observing a relevant IV can further improve the informativeness of predictive inference. See Appendix~\ref{sec:bound_obs_prob} for details.
\end{remark}

\section{Methodology}\label{sec:methodology}

The previous results show that oracle distribution-free inference is
possible when $\Gamma_P > 1-\alpha$. However, whether such inference can be attained in practice is a separate question. Specifically, while the semi-oracle prediction set $\Ch^{\mathrm{sor}}$ in~\eqref{eqn:pred_set_oracle} achieves valid distribution-free coverage,
it requires the unknown miscoverage function $\delta$ and therefore cannot be implemented directly.
In this section, we discuss practical methods that approximately attain 
coverage by estimating certain components (i.e., nuisance parameters) of the distribution. 

A straightforward approach might be to directly estimate 
the oracle miscoverage 
function $\delta(t,\cdot)$ 
with an estimator $\hat{\delta}(t,\cdot)$ constructed on training data separate from the calibration data. 
This would lead to
the prediction set defined as
\begin{equation}\label{eqn:pred_set_plugin}
    \Ch^{\mathrm{naive}}(x)=\{y\in\R:s(x,y)\le \hat{t}_\alpha\}, \text{ where } \hat{t}_\alpha=\inf\left\{t\in\R:\frac{1}{n+1}\left(\sum_{i=1}^{n}\hat{\delta}(t,X_i)+1\right)\le\alpha\right\},
\end{equation}
While this approach is reasonable, 
unfortunately its coverage gap
has only a first-order bias, of size as large as $\sup_{t \in \R} \EE{\big|\hat{\delta}(t,X) - \delta(t,X)\big|}.$ 
Specifically, Theorem~\ref{thm:plug_in} in Appendix~\ref{app:plug_in} shows that if 
the function $t\mapsto \hat{\delta}(t,x)$ is nonincreasing for all $x \in \X$ and takes values in $[0,1]$ (which is reasonable as $\delta$ has these properties), then the prediction set $\Ch^{\mathrm{naive}}(\cdot)$ in~\eqref{eqn:pred_set_plugin} satisfies
\begin{equation}\label{pl}
    \PP{Y_{n+1} \in \Ch^{\mathrm{naive}}(X_{n+1})} \geq 1-\alpha - \sup_{t \in \R} \EE{\big|\hat{\delta}(t,X) - \delta(t,X)\big|}.
    \end{equation}

This, however, is a generic approach that does not leverage the structure of the functional $\delta$.
It turns out that, through a more careful construction,
 by leveraging the structure of the functional 
 $\delta$, 
 we can develop methods that have a generally smaller converge gap.

\subsection{Doubly robust inference with efficient influence function}\label{sec:dr_eif}

Next, we provide an overview of our approach.
We start with a given score function $s:\X\times\R\to\R$, and consider prediction sets of the form $C_t(x)=\{y:s(x,y)\le t\}$. Throughout this section, we write $S=s(X,Y)$ for the corresponding score variable.

Our approach 
aims to use influence functions 
in order to derive doubly robust estimators of the coverage error $\PP{S>t}$; see e.g., 
\citet{levit1974optimality,Bickel1993,kennedy2022semiparametric} for relevant background on semi-parametric efficiency theory. 
To start with, 
a key challenge is that 
the coverage error itself is not identifiable. 
Therefore the first step is to derive an identifiable upper bound on it. This is where the functional $\delta$ appears. 
As a key step in our analysis, 
we show an upper bound in terms of the identifiable functional $\delta$:
\[\PP{S>t}\leq \EE{\delta(t,X)}, \qquad
\text{where }\delta(t,x)=1-\int_{\R} i_{P,x}(y)\,\One{s(x,y)\le t}\,\mathrm{d}y,\]
see the proof of Theorem~\ref{thm:coverage}. 
For fixed $s$ and $t$, this bound is sharp (i.e., unimprovable) over full data distributions with the same observed  distribution. See Appendix~\ref{sec:miscoverage_sharpness} for details.

However, this bound cannot be used directly
 to derive an efficient influence function (EIF) and leverage double robustness, 
because the functional
$P\mapsto\Ep{P}{\delta(t,X)}$ can fail to be pathwise differentiable due to the pointwise maximum defining $i_{P,x}$---see Theorem~\ref{thm:envelope_nonregularity} in Appendix~\ref{sec:envelope_nonregularity}.

Consequently, in the following section, we construct a function $\tilde{\delta}(\cdot,\cdot)$ that is identifiable, pathwise differentiable, and satisfies
\begin{equation}\label{eqn:tail_ineq}
    \PP{S>t}\leq \EE{\delta(t,X)}\leq \EE{\tilde{\delta}(t,X)},
\end{equation}
while remaining sufficiently close to the original bound $\delta(\cdot,\cdot)$ to ensure that the resulting procedure controls coverage as tightly as possible.

Before presenting the specific construction of the function $\tilde{\delta}$, we first describe the overall steps of the procedure and the coverage guarantee it provides. 
Consider a set of nuisance parameters $\eta$ of $P$, 
with respect to which we want to improve the sensitivity in estimation errors and do so by using 
influence functions. Assuming that the function $\tilde{\delta}$ is given, 
let $\phi_t(O;P)$ denote the EIF of the functional $P\mapsto\Psi_t(P)=\Ep{P}{\tilde{\delta}(t,X)}$ with respect to the nuisance functions $\eta$, and let $\psi_t(O;P)=\Psi_t(P)+\phi_t(O;P)$ denote its uncentered version. In the specific construction that will be introduced, $\psi_t(O;P)$ depends on $P$ only through the nuisance functions $\eta$, so we write it also as $\psi_t(O;\eta)$. Then, for any nuisance estimator $\hat{\eta}$,
\[\Ep{P}{\psi_t(O;\hat{\eta})} = \Ep{P}{\psi_t(O;\eta)} + R_t(\hat{\eta},\eta) = \Psi_t(P) + R_t(\hat{\eta},\eta),\]
where the remainder term $R_t(\hat{\eta},\eta)$ is of second order in the nuisance estimation error (see, e.g.,~\citet{robins2017minimax}). For the construction that will be introduced,
the remainder term is not only second order but also exhibits double robustness, 
in the sense that it consists of products of nuisance estimation errors.

Now we consider the following prediction set, where instead of 
$\hat \delta$,  we use the efficient influence function $\psi_t$ in the constraint:
\begin{equation}\label{eqn:pred_set_dr}
    \Ch(x) = \big\{y\in\R:s(x,y)\leq \hat{t}_\alpha\big\}, \quad \text{ where }\; \hat{t}_\alpha = \sup\left\{t\in\R : \frac{1}{n}\sum_{i=1}^n \psi_t(O_i;\hat{\eta}) \geq \alpha\right\}.
\end{equation}
We obtain the following coverage guarantee for the prediction set~\eqref{eqn:pred_set_dr}.

\begin{lemma}\label{lem:coverage_dr}
    Suppose $L\leq\psi_t(O;\hat\eta)\leq U$ holds almost surely for every $t\in\R$, for constants $L<U$. Suppose also that~\eqref{eqn:tail_ineq} holds.
    Then the prediction set $\Ch(\cdot)$ in~\eqref{eqn:pred_set_dr} satisfies
    \[\PP{Y_{n+1} \in \ch(X_{n+1})} \geq 1-\alpha - \sup_{t \in \R} |R_t(\hat{\eta},\eta)| - (U-L)\sqrt{\frac{\pi}{8n}}.\]
\end{lemma}
 We expect the error term  
$\sup_{t \in \R} |R_t(\hat{\eta},\eta)|$
 in this lower bound to be smaller than the one for the naive plug-in method as the residual is of second order; whereas the second term has a parametric $n^{-1/2}$ error rate which is also expected to be of smaller order than the generically nonparametric first error rate. 

Deriving the above bound on the coverage
is more challenging than bounding the estimation error of a fixed tail probability $\PP{S > t}$, as the prediction set involves a random threshold $\hat{t}$.
The proof of Lemma~\ref{lem:coverage_dr} addresses this challenge through a careful analysis that introduces several deterministic approximations of $\hat{t}_\alpha$, while leveraging the monotonicity of $t \mapsto \PP{S > t}$ and avoiding union bounds that can lead to conservative bounds.

\subsection{Policy-based inference}\label{sec:method_policy}

Now we introduce our procedure. 
Throughout this section, we use the following notation for the
values of the nuisance functions: $\kappa$ for the feature-conditional distribution of 
the IV, and $q$ for the
retained portion for fixed instrument from \eqref{n1}, which is maximized over the instrument level in $i_P$: 
\[\kappa_{z,x}=\PPst{Z=z}{X=x},\qquad
q_{z,x}(y)=\pi_{z,x}f_{z,x,1}(y),
\quad \text{for } z\in\Z \text{ and } x\in\X.\]
For an IV distribution $P \in \Piv$, we will often write $\eta_P$ to denote the collection of nuisance 
parameters corresponding to $P$, 
i.e., $\eta_P = (\kappa_{z,x}, q_{z,x})_{z \in \Z,x \in \X}$.

To address the non-differentiability of the functional, instead of taking the maximum over all values of the instrument, we consider a \emph{policy} $a:\X\times\R\rightarrow\Z$ that selects an instrument $z\in\Z$ for each $(x,y)$. We learn the policy on a separate dataset to select informative instruments, and then use an EIF-based approach to conduct inference for the learned policy.\footnote{This separation of policy learning and evaluation is analogous to the approach used in the literature on individualized treatment rules~\citep{vanderlaan2015targeted,luedtke2016optimal}.}

Each policy provides an upper bound on $\delta$, since
for any $x\in\X$ and $y\in\R$, 
\[q_{a(x,y),x}(y)\leq \max_{z\in\Z} q_{z,x}(y)=i_{P,x}(y).
\]
 Therefore, the function $\tilde{\delta}_a:\R\times\X\rightarrow\R$ defined by
\begin{equation}\label{eqn:delta_policy}
    \tilde{\delta}_a(t,x)=1-\int_{\R} q_{a(x,y),x}(y)\,\One{s(x,y)\leq t}\,\mathrm{d}y
\end{equation}
is a valid upper bound on $\delta$, 
which we can use as an input to our approach from
Section~\ref{sec:dr_eif}. 

We now derive the efficient influence function of the bound $\Ep{P}{\tilde{\delta}_z(t,X)}$. Let us define the following pseudo-outcome, referred to as the \emph{pseudo-coverage indicator}, and its conditional mean:
\[ H^a_{z,t}(O)=R\cdot \One{S \leq t}\One{a(X,Y)=z}, \text{ and }  m^a_{z,x}(t)=\mathbb E\{H^a_{z,t}(O)\mid X=x,Z=z\},\]
for $t \in \R, x \in \X$, and $z \in \Z$.
Then for nuisance functions $\eta=(\kappa_{z,x},m_{z,x}^a)_{z\in\Z, x \in \X}$, 
the (uncentered) EIF of $P\mapsto \Ep{P}{\tilde{\delta}_a(t,X)}$
in the nonparametric observed data model is
\begin{equation}\label{eqn:policy_eif}
\psi_{a,t}(O;\eta)=1-\sum_{z \in \Z}\left[m_{z,X}^a(t)+\frac{\One{Z=z}}{\kappa_{z,X}}\{H_{z,t}^a(O)-m_{z,X}^a(t)\}\right].
\end{equation}

 We construct nuisance estimators 
$\hat{\kappa}_{z,x}$ and $\hat{m}_{z,x}$ using separate training data, 
for instance using the same data used to construct the non-conformity score $s$, and obtain an estimated $\hat \eta$.
Then we define the prediction set as
\begin{equation}\label{eqn:pred_set_policy}
    \Ch(x) = \big\{y\in\R:s(x,y)\leq \hat{t}_\alpha\big\}, \quad \text{ where }\; \hat{t}_\alpha = \sup\left\{t\in\R : \frac{1}{n}\sum_{i=1}^n \psi_{a,t}(O_i;\hat{\eta}) \geq \alpha\right\}.
\end{equation}
Applying Lemma~\ref{lem:coverage_dr}, 
 we obtain the following result for the coverage of the above  prediction set. 

\begin{theorem}\label{thm:policy_coverage}
Fix any policy $a:\X\times\R\rightarrow\Z$. Suppose $0\le\hat m_{z,X}^a(t)\le1$ and $b\le\hat\kappa_{z,X}\le1$ almost surely for every $z\in\Z$ and $t\in\R$, for some $b>0$.  Then the prediction set $\Ch(\cdot)$ in~\eqref{eqn:pred_set_policy} satisfies
\[\PP{Y_{n+1}\in\Ch(X_{n+1})}\geq 1-\alpha -\frac{1}{b}\sum_{z\in\Z}\|\hat{\kappa}_{z,X}-\kappa_{z,X}\|\cdot\sup_{t\in\R}\|\hat{m}_{z,X}^a(t)-m_{z,X}^a(t)\|-C\sqrt{\frac{1}{n}},
\]
where $C$ depends only on $K$ and $b$, and $\|\cdot\|$ denotes the $L^2(P_X)$-norm.
\end{theorem}

The boundedness condition on the estimator $\hat{\kappa}_{z,x}$ can be satisfied by simple clipping---although this, of course, implies that in extreme settings where the true values of $\kappa_{z,x}$ are very close to zero, we might obtain a nontrivial coverage gap.

\begin{remark}
    In Section~\ref{sec:upper_bound} of the Appendix, we also derive an upper bound on the coverage rate of the prediction set~\eqref{eqn:pred_set_policy}. Specifically, we establish an upper bound of the form $1-\alpha+(1-\Gamma_P)+\text{(error terms)}$, indicating that the tightness of the procedure depends significantly on the identification index $\Gamma_P$. In the experiments that will be presented, we indeed observe that the coverage rate is more tightly controlled for larger values of $\Gamma_P$. 
\end{remark}

\begin{remark}
When the instrument propensity $P_{Z\mid X}$ is known, an alternative procedure based on weighted conformal prediction can achieve exact finite-sample coverage. However, this approach can be overly conservative and less practical due to inefficient use of the available samples and conservative estimation of the identification index, with the lack of double robustness. See Appendix~\ref{sec:wcp_known_pzx} for further details.
\end{remark}

\subsubsection{Choice of the policy}\label{sec:policy_choice}

We now describe one approach to choosing a policy that yields a tight approximation, $\EE{\delta(t,X)} \leq \EE{\tilde{\delta}_a(t,X)}$, 
thereby allowing tight coverage control.

 At the population level, 
the pointwise optimal policy is given by
\begin{equation}\label{eqn:policy_optimal}
    a^*(x,y)\in\argmax_{z\in\mathcal Z}q_{z,x}(y),
\end{equation}
where $q_{z,x}(y)=\pi_{z,x}f_{z,x,1}(y)$. We aim to imitate this oracle policy at the empirical level.
 
Suppose that the nuisance parameters are estimated using a separate training data $(O_i^\text{tr})_{i \in [n_{\mathrm{train}}]} = \left(X_i^{\mathrm{tr}},Z_i^{\mathrm{tr}},R_i^{\mathrm{tr}},R_i^{\mathrm{tr}}Y_i^{\mathrm{tr}}\right)_{i \in [n_{\mathrm{train}}]}$.
 Of course in order to ensure accuracy, these  
 would ideally 
 be sampled from a distribution that shares the necessary nuisance functions.
We first estimate the  densities $f_{z,x,1}$ and $\pi_{z,x}$ using the training data and define
\(
\hat q_{z,x}(y)
=
\hat\pi_{z,x}\hat f_{z,x,1}(y).
\)

Then,
let $\mathcal A$ denote a prespecified class of candidate policies,  e.g., parametrized by random forests, neural nets, etc.
For each observed training outcome, we can then construct a surrogate label
\[
L_i
=
\argmax_{z\in\mathcal Z} 
\hat q_{z,X_i^{\mathrm{tr}}}(Y_i^{\mathrm{tr}}),
\qquad i\in\{j:R_j^{\mathrm{tr}}=1\},
\]
and subsequently fit a classifier from the prespecified policy class $\mathcal A$ using these observed $(X_i^{\mathrm{tr}},Y_i^{\mathrm{tr}})$ pairs as predictors and $L_i$ as the response. The overall procedure, including this policy selection step, is summarized in Algorithm~\ref{alg:policy}.

\begin{algorithm}[ht]
\caption{Policy-based predictive inference with IV}
\label{alg:policy}
\KwIn{Training data $(O_i^\mathrm{tr})_{i \in [n_\mathrm{train}]}$, Calibration data $(O_i)_{i \in [n]}$, Test input $X_{n+1}$, Policy class $\mathcal A$, Target level $\alpha$.}

\KwOut{Prediction set $\Ch(X_{n+1})$.}

Construct a nonconformity score $s:\X\times\R\rightarrow\R$ using the training data.

Construct estimators $\hat\pi_{z,x}$ and $\hat f_{z,x,1}(y)$ using the training data, and define
\(
\hat q_{z,x}(y)=\hat\pi_{z,x}\hat f_{z,x,1}(y)\), \(z\in\Z.
\)

For each $i$ such that $R_i^{\mathrm{tr}}=1$, construct the surrogate label
\(
L_i=\argmax_{z\in\Z}\hat q_{z,X_i^{\mathrm{tr}}}(Y_i^{\mathrm{tr}}).
\)

Learn a policy $\hat a\in\mathcal A$ by fitting a classifier using $(X_i^{\mathrm{tr}},Y_i^{\mathrm{tr}})$ as predictors and $L_i$ as the response, over $i$ such that $R_i^{\mathrm{tr}}=1$.

Construct the nuisance estimators $\hat\eta=(\hat \kappa_{z,x},\hat m_{z,x}^{\hat a})_{z\in\Z}$ using the training data.

Compute
\(
\hat t_\alpha=\sup\left\{t\in\R:\frac1n\sum_{i=1}^n\psi_{\hat a,t}(O_i;\hat\eta)\ge\alpha\right\},
\)
where $\psi_{\hat a,t}$ is defined according to~\eqref{eqn:policy_eif}.

\Return Prediction set $\Ch(X_{n+1})=\{y\in\R:s(X_{n+1},y)\le\hat t_\alpha\}$.

\end{algorithm}

\subsection{Remarks on the efficiency of the procedure}\label{sec:rmk}
A natural question for a predictive inference procedure is how to define and measure its efficiency. This is generally difficult in the distribution-free setting. One may consider empirically investigating how tightly the coverage rate is controlled, but this can vary across settings---a distribution-free procedure controls the worst-case coverage over all distributions. 

However, a basic consideration of
the coverage--width tradeoff
attained by the proposed procedure
may provide some insight.
Recall that our prediction sets have the form $\{y : s(x,y) \leq \hat{t}\}$, as in standard conformal prediction.
The only difference is
how the score bound $\hat{t}$ is chosen to account for the missingness pattern.
Therefore, as the target level $\alpha$ varies, the proposed procedure exhibits a coverage-width tradeoff analogous to that of standard conformal prediction, with comparable efficiency.
For instance,
suppose we use the standard residual score $s(x,y) = |y - \hat{\mu}(x)|$. Then the width of the prediction set is exactly $2 \cdot \hat{t}$, implying that the same width corresponds to the same prediction set and hence the same (threshold-)conditional coverage rate. This intuition generally extends to other choices of the score function.

\section{Simulations}\label{sec:simulation}

\subsection{Coverage evaluation}\label{sec:sim_cov}

Here, we provide simulation results to illustrate the coverage properties of the proposed procedure.

\subsubsection{Data generating process}\label{sec:dgp}

We generate the data according to the following distribution.
 For some dimension $p$, we generate the features 
 $X\sim \mathcal{N}(0,I_p)$ and sample
 a categorical instrumental variable whose distribution is determined by the alignment of some parameters $\theta_z$ with the features 
 $x$:
\begin{align*}
&Z\mid X=x\sim\operatorname{Categorical}\big((\kappa_{z,x})_{z\in\mathcal Z}\big),\qquad \kappa_{z,x}:=\frac{b_z\exp(a\theta_z^\top x)}{\sum_{k\in\mathcal Z}b_k\exp(a\theta_k^\top x)}.
\end{align*}
We then sample the outcomes given the features and instrumental variables as a mixture of two Gaussian distributions with different shifts:
\begin{align*}
&Y\mid X=x,Z=z\sim q\cdot \mathcal{N}(\mu(x)+c_1,\sigma^2)+(1-q)\cdot \mathcal{N}(\mu(x)+c_2,\sigma^2)\;\;\forall z,\qquad \mu(x):=\beta^\top x.
\end{align*}
This mixture retains a simple linear dependence on the features while allowing non-Gaussian outcome distributions, including the asymmetric and symmetric bimodal distributions in Settings 1 and 2 below. In both settings, $qc_1+(1-q)c_2=0$, so $\mu(x)$ remains the conditional mean. This construction allows us to evaluate coverage under misspecification of the Gaussian working model used below to estimate the observed-outcome density $f_{z,x,1}$. Using the same conditional outcome distribution for every $z$ ensures the IV validity condition $Y\indep Z\mid X$ in Assumption~\ref{asm:iv}.

 Finally we set the missingness probability as a function of a
 hyperparameter $\Gamma$ as follows: 
\begin{align*}
&R\mid X=x,Z=z,Y=y\sim\operatorname{Bernoulli}\left(\pi_{z,x}+\rho\frac{\pi_{z,x}}{\Gamma}(\Gamma-\pi_{z,x})\{h_x(y)-\bar h\}\right),
\end{align*}
where 
\[
\pi_{z,x}:=\Gamma \frac{\kappa_{z,x}}{\max_{k\in\mathcal Z} \kappa_{k,x}},\quad h_x(y):=2\One{y>\mu(x)}-1, \quad \bar h:=\EEst{h_X(Y)}{X}.
\]
In Appendix~\ref{sec:dgp_id}, we show that the identification index $\Gamma_P$ of the above distribution is exactly equal to the parameter $\Gamma$, regardless of the choice of other parameters. Therefore, in the simulations, we can directly control the identification index without empirically estimating its value.
 For this simulation, we choose 
$p=10, \mathcal Z=\{1,\ldots,10\}$, and $\beta=(1,0.8,0.6,0.4,0.2,-0.2,-0.4,-0.6,-0.8,-1)^\top$.

We consider two settings that illustrate 
different directions of coverage error
under misspecification of the outcome model. Setting 1 uses an asymmetric outcome mixture with a dominant component and a small, distant component, whereas Setting 2 uses a symmetric mixture of equally weighted separated components.
As shown below, the plug-in method often over-covers in Setting 1 and under-covers in Setting 2, while the policy-based method generally yields coverage closer to the semi-oracle benchmark in both settings.

\paragraph{Setting 1.}
We set
\[(a,q,c_1,c_2,\sigma,\rho,\bar h)=(1,0.9,-1,9,0.25,0.2,-0.8),\qquad (b_1,b_2,\ldots,b_{10})=(0.4,0.6/9,\ldots,0.6/9),\]
and
\[\theta_1=(0.16,-0.11,0.08,0,\ldots,0)^\top,\quad \theta_z=\frac{z-1}{9}(-0.05,0.04,-0.03,0,\ldots,0)^\top,\quad z=2,\ldots,10.\]

\paragraph{Setting 2.}
We set
\[(a,q,c_1,c_2,\sigma,\rho,\bar h)=(0.5,0.5,-2.5,2.5,0.25,0.8,0),\quad b_z=1,\quad z=1,\ldots,10,\]
with $\theta_z$ constructed using equally spaced values between $-1.5$ and $1.5$.

\subsubsection{Results}

We generate a training dataset of size $n_\text{train}\in\{500,2000\}$ by sampling from the true data-generating distribution
and estimate the nuisance 
parameters $\kappa_{z,x}=\PPst{Z=z}{X=x}$, $\pi_{z,x}=\PPst{R=1}{X=x,Z=z}$, and $f_{z,x,1}$. 
We use multinomial logistic regression 
to estimate 
$\kappa_{z,x}$, random forests for $\pi_{z,x}$. 
We fit a Gaussian density for $f_{z,x,1}$ 
by estimating the required conditional mean and variance, this allows us to evaluate the performance of our method in a misspecified setting.
The policy is learned using the procedure described in Section~\ref{sec:policy_choice} with random forest classification, and we use the score $s(x,y) = |y - \hat{\mu}(x)|$ where $\hat{\mu}$ is constructed using random forest regression. We then conduct 200 independent trials, generating calibration and test datasets of size $500$ and applying the procedures at levels $\alpha=0.05,0.1,\ldots,0.3$. Specifically, we apply the following procedures:
\begin{enumerate}
    \item[(i)] \textit{Semi-oracle}: Prediction set $\Ch^{\mathrm{sor}}$ from~\eqref{eqn:pred_set_oracle}.
    \item[(ii)] \textit{Plug-in}: Prediction set $\Ch^{\mathrm{naive}}$ from~\eqref{eqn:pred_set_plugin}.
    \item[(iii)] \textit{EIF-oracle}: Procedure~\eqref{eqn:pred_set_policy} with the true nuisance functions $\eta$ in place of $\hat{\eta}$, and with the optimal policy~\eqref{eqn:policy_optimal}.
    \item[(iv)] \textit{Policy-based}: Procedure~\eqref{eqn:pred_set_policy} with the estimated nuisance functions $\hat{\eta}$ and the policy constructed using training data.
\end{enumerate}
The semi-oracle and EIF-oracle methods (i) and (iii) are both sample-based approximations of the idealized oracle prediction set $\Ch^{\mathrm{or}}$ in~\eqref{eqn:pred_set_oracle_2}. We repeat this experiment 
for three values of the identification index, $\Gamma=0.8,0.9,1$.

\begin{figure}[htbp]
    \centering
    \includegraphics[width=0.9\linewidth]{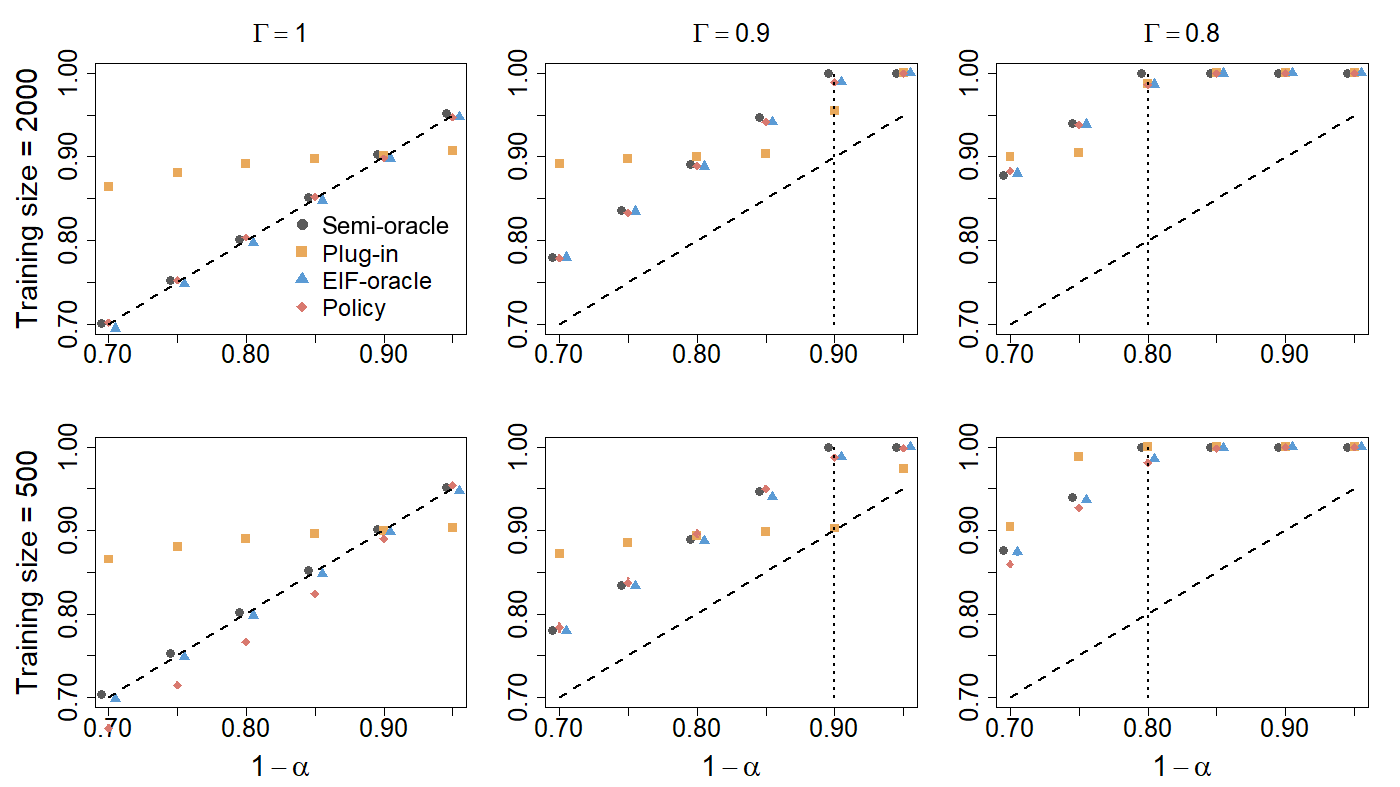}
    \caption{Coverage rates of the prediction set from Procedure~\ref{alg:policy} and the plug-in method, along with their oracle counterparts, under identification index $\Gamma = 0.8,0.9,1$, with error bars, in Setting 1. The dotted lines represent the $y=x$ and $x=\Gamma$ lines.}
    \label{fig:coverage_policy_1}
\end{figure}

\begin{figure}[htbp]
    \centering
    \includegraphics[width=0.9\linewidth]{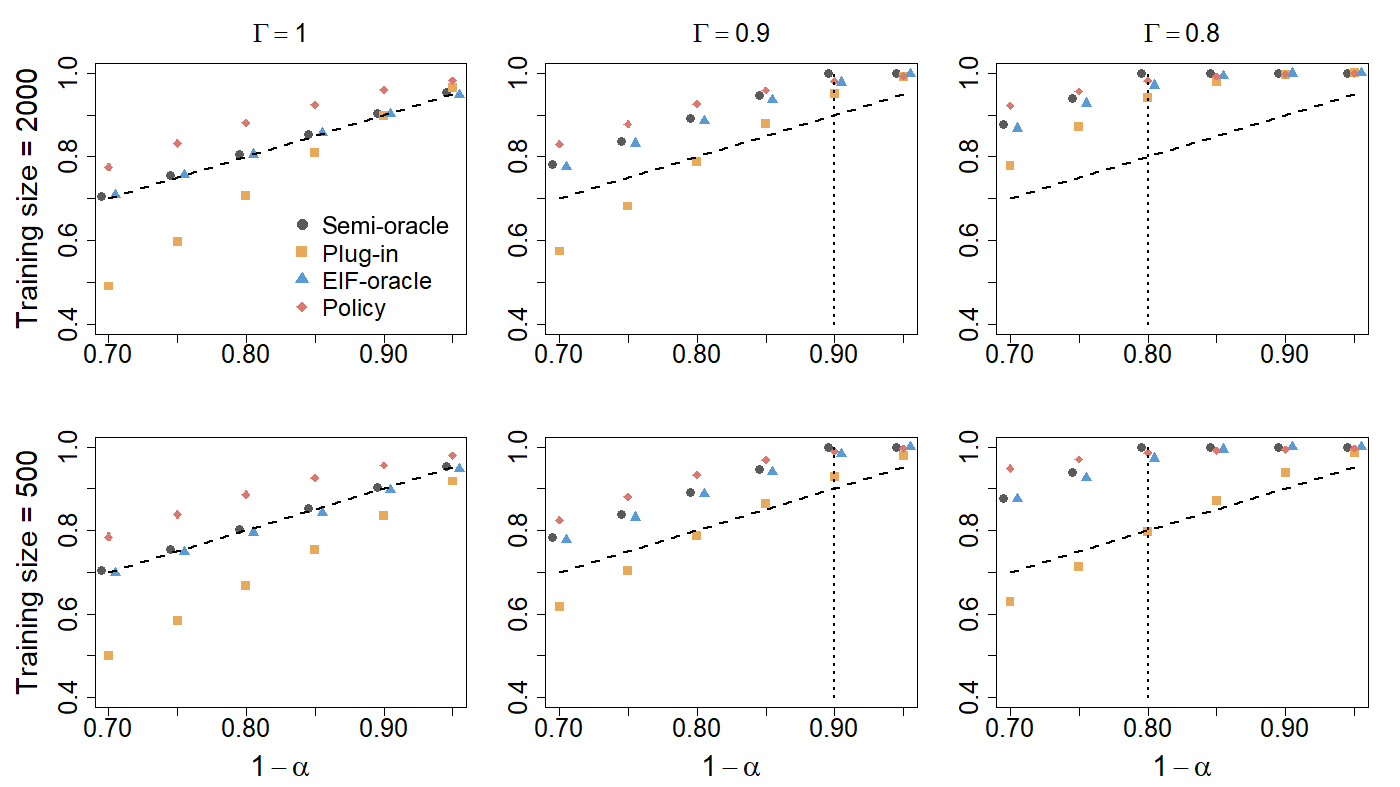}
    \caption{Coverage rates of the prediction set from Procedure~\ref{alg:policy} and the plug-in method, along with their oracle counterparts, under identification index $\Gamma = 0.8,0.9,1$, with error bars, in Setting 2. The dotted lines represent the $y=x$ and $x=\Gamma$ lines.}
    \label{fig:coverage_policy_2}
\end{figure}

Figures~\ref{fig:coverage_policy_1} and~\ref{fig:coverage_policy_2} show the coverage rates of the various procedures. 
The semi-oracle and EIF-oracle methods show nearly identical performance, and both achieve the desired level of coverage. When $\Gamma=1$, they achieve tight 
control, and as $\Gamma$ decreases---i.e., the unidentified portion of the distribution increases---the coverage control tends to become less tight. 
When $\Gamma<1-\alpha$, the coverage of these methods
is close to unity, 
consistent with the theoretical result that nontrivial prediction sets cannot be obtained when $\Gamma<1-\alpha$ even by oracle methods. 

The plug-in procedure shows relatively unstable behavior and fails to closely approximate the semi-oracle prediction set $\Ch^{\mathrm{sor}}$ in many cases. As a result, it often exhibits substantial over-coverage (Setting 1) or under-coverage (Setting 2). On the other hand, 
our policy-based method overall closely approximates $\Ch^{\mathrm{sor}}$ and consequently maintains coverage around or above the target level.

While the 
plug-in method can achieve tighter coverage control in a few settings---e.g., when the semi-oracle and EIF-oracle methods are conservative due to a low identification index---it can be difficult in practice to 
know which regime we are in, as we do not have the ground truth labels.
Based on both the theoretical discussion and the empirical results, we therefore generally recommend the EIF-based policy method over the plug-in method.

\subsection{Comparison with weighted conformal prediction}\label{sec:wcp}

To our knowledge, conformal prediction has received little attention in the IV setting, and thus there are no direct comparison methods for our approach. Still, one may consider comparisons with methods that one may perhaps naively use in our setting. 
In this regard, we include a comparison with weighted conformal prediction (WCP)---which does \emph{not} make use of the instrumental variable and instead assumes that the outcomes are missing at random.
 To be clear, our
 goal here is not to show that WCP would fail, because it is clearly not designed for our setting,
 but rather to show our best effort
 comparison with a (naive) baseline. 

Like our procedure, 
weighted conformal prediction requires estimating 
nuisance functions;
specifically the propensity score.
The performance of both methods depends on the quality of these estimators. 
In this experiment, we compare the oracle versions of the two methods, which use the true nuisance functions, optimal policy, and propensity score, respectively. We 
do this to focus on the properties of the inference procedures themselves, without confounding the comparison with estimation error in the nuisance functions.

We conduct the experiments under three scenarios for the data-generating distribution:
\begin{enumerate}
\item Setting 1: The identification index is close to unity, and our proposed method is expected to perform well.
\item Setting 2: The identification index is smaller, and the proposed method is likely to over-cover, while the outcome remains MNAR, making WCP unreliable.
\item Setting 3: The two distributions $P_{Y \mid X, R=1}$ and $P_{Y \mid X}$ are nearly identical, making WCP likely to perform well.
\end{enumerate}
More details on the data-generating distributions for these settings are provided in Appendix~\ref{sec:method_detail}.

\begin{figure}
    \centering
    \includegraphics[width=0.9\linewidth]{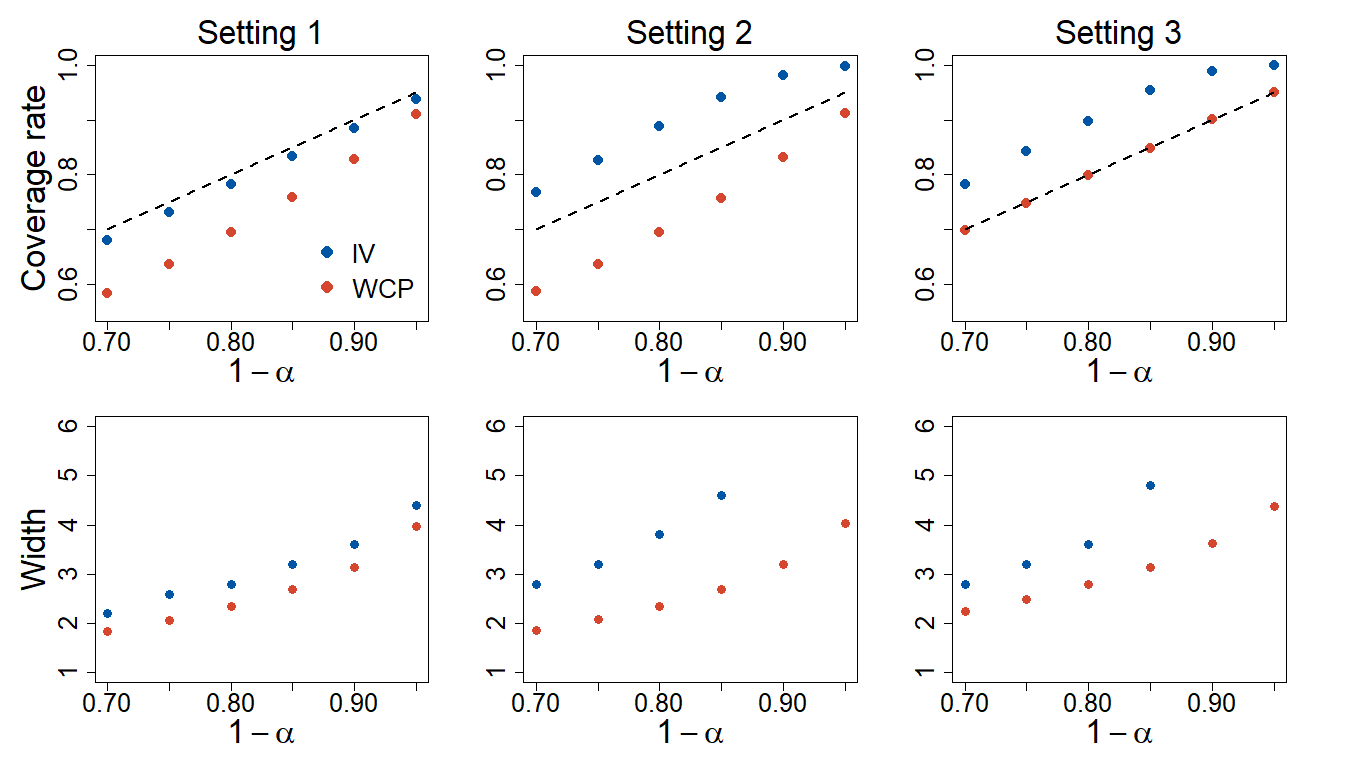}
    \caption{Coverage rate and median width of the prediction set from Procedure~\ref{alg:policy} and weighted conformal prediction across different settings. Cases with infinite width are not plotted.}
    \label{fig:wcp}
\end{figure}

In all settings, we repeat the experiment 200 times, generating calibration and test data, each of size 500, applying Procedure~\ref{alg:policy} and WCP to the calibration data, and computing the empirical coverage and width of the resulting prediction sets. The results are shown in Figure~\ref{fig:wcp}.

The results illustrate that the proposed procedure controls the coverage rate around or above the target level $1-\alpha$ regardless of the underlying distribution, even when this requires the prediction interval to be overly wide.
This is consistent with its distribution-free coverage guarantee. In contrast, weighted conformal prediction fails to control the coverage rate,
except in the special case of setting 3 where the distribution shift is primarily in the covariate.

\section{Experiments with empirical data}
\label{sec:empirical-data}

We illustrate the methods using data from the UK Family Expenditure Survey (FES) and from the Third National Health and Nutrition Examination Survey (NHANES III). 
Evaluation of the procedure based on empirical data is challenging as the missing 
outcomes are not observed and so the actual coverage cannot be directly evaluated. 
For example, evaluating coverage only among individuals with observed outcomes would assess coverage conditional on $R=1$, rather than marginal coverage, which can differ from
our target
under nonrandom missingness.
To deal with this challenge, 
we conduct the following two approaches:
\begin{enumerate}
    \item \textbf{Semi-synthetic experiment}: We use only datapoints with observed outcomes from the original datasets. We retain their observed $X$ and $Y$, construct $Z$ from the recorded instrument, and generate the missingness indicator $R$ via simulation. 
    This enables us to empirically evaluate the coverage rate of the procedure.
    
    \item \textbf{Full empirical experiment}: We retain the original missingness pattern and interpret the prediction interval widths relative to the observed outcome distributions, to illustrate how much information the procedure provides about individual outcomes (though without being able to evaluate coverage).
\end{enumerate}

\subsection{Datasets and methods}

\subsubsection{UK Family Expenditure Survey (FES)}

Using FES data from 1978 to 2000, \citet{arellano2017} studied changes in UK wage inequality while accounting for selection into employment. A key challenge is that wages are observed only for individuals who are employed, and thus the observed wage distribution may differ from the wage distribution of the broader population. 
They use potential out-of-work benefit income as an instrument, as it is related to the decision to work,
but is assumed to be conditionally independent of wages given the observed covariates. We use their replication data to illustrate our method.
The goal is to construct
prediction sets for the unobserved wages of nonworkers.

We consider single men aged 23--59, with log hourly wage as the outcome $Y$. The covariates $X$ include age, schooling, survey year, region, and the number of children in each of six age groups. We discretize log potential benefit income into tertiles to define a discrete instrument. After applying the eligibility restrictions and selecting one 
person per household, the resulting cohort contains 18,156 men, of whom 13,226 have observed wages. 

\subsubsection{Third National Health and Nutrition Examination Survey (NHANES III)}

NHANES III contains health and nutritional information collected in the US from 1988 to 1994. We use reported height as the outcome $Y$ \citep{nhanes3}. The covariates $X$ include age, education and its missingness indicator, sex, race/ethnicity, education status, interview language, census region, broad urbanization, and survey phase.

We use adult household interviewer identity as a candidate instrument. 
Specific interviewers
may affect the chances that a participant reports their height, whereas the participant's underlying height is determined before the interview. Adjustment for demographic characteristics, geography, and other features is intended to account for factors associated with both interviewer assignment and height. 
This motivates the IV condition $Z \indep Y \mid X$.

Using an independent 20\% family design sample, which is excluded from the subsequent analysis, we select the five most frequent interviewers and retain their identities as the five categories of $Z$. After sampling one adult per family, the resulting analysis cohort contains 2,144 adults, including 2,052 with observed height.

\subsubsection{Methods}

We apply Procedure~\ref{alg:policy} using the standard absolute residual score $s(x,y)=|y-\hat{\mu}(x)|$, where $\hat{\mu}$ is obtained by
applying random forest regression to a training split.
The nuisance functions are estimated on the training split using multinomial logistic regression for the instrument propensities $\kappa_{z,x}$, random forests for the response probabilities $\pi_{z,x}$, and Gaussian working models for $Y\mid X,Z,R=1$, with conditional moments estimated using random forests. Outcomes are standardized using the observed training sample moments for model fitting. 
For comparison, we consider weighted conformal prediction, using either $X$ or $(X,Z)$ as covariates, denoted as $X$-WCP and $(X,Z)$-WCP.

\subsection{Semi-synthetic experiment with added missingness}

Due to the challenges of evaluation explained earlier, 
we first conduct a semi-synthetic experiment, where we use the datapoints $(X,Y,Z)$ from the fully observed datapoints and generate the missingness $R$. We sample one person per household or family. We split the data into three disjoint subsets: a pilot sample, a training sample, and an evaluation pool. The three datapoints have sizes $(1000, 2000, 2000)$ for FES and $(208, 843, 1031)$ for NHANES III. For NHANES III, we construct the sample used for the semi-synthetic experiment by selecting one participant with observed height from each of 2,082 families. The pilot sample is used to fit a prediction model $\hat{\mu}_0$ and estimate its out-of-bag residual standard deviation $\tau$, which are used to induce additional missingness. 
The training sample is used to fit all nuisance functions required by the procedures, with the fitted models held fixed across trials. The evaluation pool is used to repeatedly draw calibration and test datapoints to evaluate the empirical coverage and width of the procedures.

 We repeat the experiment over 200 trials, drawing disjoint calibration and test datapoints of size 500 each from the evaluation pool for both datasets, generating missingness indicators for the calibration sample in each trial, applying the procedures, and evaluating empirical coverage on the test sample.

We generate the missingness as follows. Let $r_z(x,y)=\PPst{R=1}{X=x,Y=y,Z=z}$ denote the conditional observation probability, and let $v=(y-\hat{\mu}_0(x))/\tau$, where the function $\hat{\mu}_0$ is constructed based on the pilot sample, and $\tau$ denotes its out-of-bag residual standard deviation. We consider two MNAR settings with a hyperparameter $g\in\{0.90,0.95,0.99,1\}$ that will control the identification index. 
In Setting 1,
\[
r_z(x,y)=
\begin{cases}
g, & z=z_\star,\\
g\cdot(0.10+0.65\cdot c_z\exp(-0.75v^2)), & z\ne z_\star,
\end{cases}
\]
where $z_\star$ is the most frequent training category, and the $c_z$ values for $z\ne z_\star$ are distinct and equally spaced over $[0.4,1]$. In Setting 2,
\[
r_z(x,y)=g\frac{h_z(v)}{\max_j h_j(v)},\qquad
h_z(v)=0.1+0.9\exp\left\{-\frac12\left(\frac{v-b_z}{0.8}\right)^2\right\},
\]
where $b_z=\Phi^{-1}((z+1/2)/K)$ for $z=0,\ldots,K-1$, with $K$ denoting the number of IV categories. Here $K=3$ for FES and $K=5$ for NHANES III.

In both settings, $\max_z r_z(x,y)=g$, and hence $\Gamma_P=g$ under the IV validity assumption, allowing us to directly control the identification index through $g$. The two settings differ in how information is distributed across instrument categories. In Setting 1, the same instrument category attains the maximum response probability for all outcome values. In Setting 2, the maximizing category varies with the outcome, so different instrument categories provide information about different regions of the outcome distribution.

\begin{figure}
\centering
\includegraphics[width=0.85\linewidth]{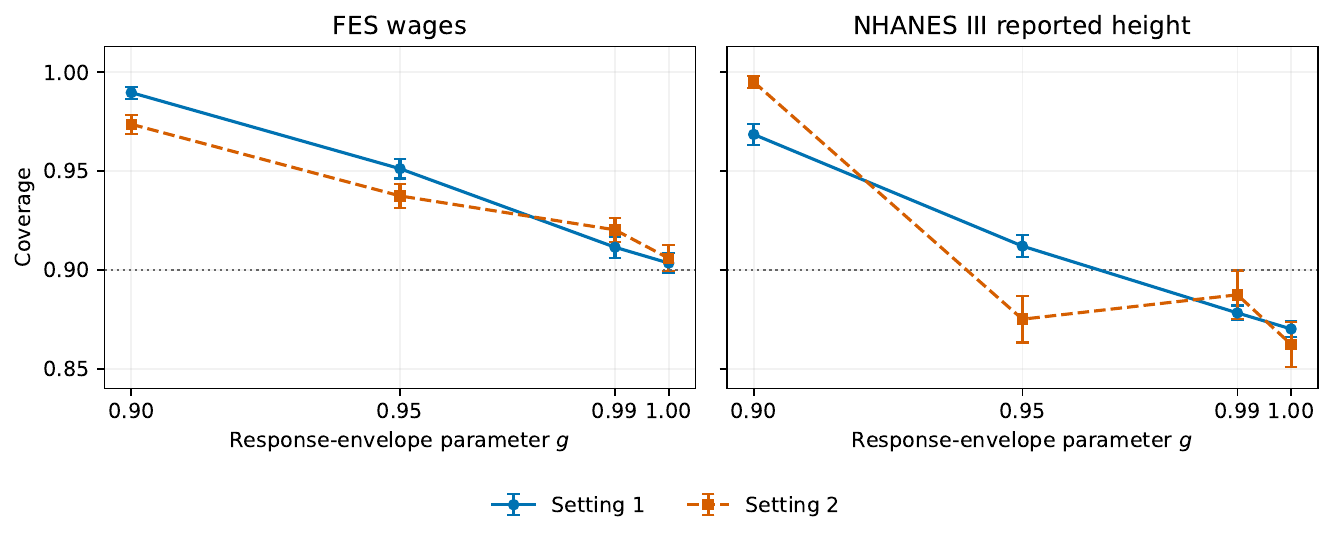}
\caption{Semi-synthetic experiments with FES and NHANES III: coverage of Procedure~\ref{alg:policy} at $\alpha=0.1$ in Settings 1 and 2, across different values of $g$, with error bars. The dotted line indicates the $0.9$ target coverage.}
\label{fig:empirical-masking}
\end{figure}

The results are shown in Figure~\ref{fig:empirical-masking}. For FES, coverage remains above the target level $1-\alpha=0.9$ across all settings considered, with tighter coverage control as $g$ increases, which is consistent with the simulation results. For NHANES III, coverage similarly approaches the target level as $g$ increases, and the procedure exhibits modest undercoverage at the largest values of $g$. Table~\ref{tab:empirical-masking} additionally reports the results for Procedure~\ref{alg:policy} and the comparison methods $X$-WCP and $(X,Z)$-WCP in Setting 1. For the FES data, the proposed procedure attains the target coverage level of $0.9$, while both weighted conformal prediction methods suffer from undercoverage. 
For the NHANES data, the proposed procedure achieves coverage close to the target level, with slight undercoverage, while the comparison methods tend to have somewhat lower coverage.

\begin{table}
\centering\small
\setlength{\tabcolsep}{4pt}
\begin{tabular}{lrrrr}
\toprule
& \multicolumn{2}{c}{FES} & \multicolumn{2}{c}{NHANES III}\\
\cmidrule(lr){2-3}\cmidrule(lr){4-5}
Method & Coverage & Mean width & Coverage & Mean width \\
\midrule
IV
& \cellcolor{green!20}0.903 (0.003)
& 1.42 (0.014)
& \cellcolor{red!20}{0.870 (0.002)}
& 8.39 (0.037) \\

$X$-WCP
& \cellcolor{red!20}0.868 (0.002)
& 1.22 (0.006)
& \cellcolor{red!20}{0.847 (0.002)}
& 7.89 (0.025) \\

$(X,Z)$-WCP
& \cellcolor{red!20}0.849 (0.002)
& 1.15 (0.006)
& \cellcolor{red!20}{0.842 (0.002)}
& 7.80 (0.034) \\
\bottomrule
\end{tabular}
\caption{Semi-synthetic experiments with FES and NHANES III: coverage rate and mean prediction interval width of Procedure~\ref{alg:policy} (IV), $X$-WCP, and $(X,Z)$-WCP, with Monte Carlo standard errors, at level $\alpha=0.1$ and parameter $g=1$ in Setting 1. Widths are in log hourly wages for FES and reported height in inches for NHANES III.}
\label{tab:empirical-masking}
\end{table}

In Section~\ref{sec:semi_synthetic_mar} of the Appendix, we additionally provide results for the MAR setting, where the instrumental variable is less informative and the WCP methods are expected to perform reasonably well.

\subsection{Full empirical experiment}
\label{sec:empirical-application}

We now apply Procedure~\ref{alg:policy} to the original datasets at level $\alpha=0.1$. We follow the same overall steps as in the semi-synthetic experiments, but use the original data with its observed missingness indicators. The FES dataset includes 2,000 training participants and 700 calibration participants, while the NHANES III dataset includes 1,072 and 536, respectively. We construct prediction intervals for 1,213 FES and 23 NHANES III participants with missing outcomes using separate held-out pools. We report the results for the proposed method below and defer the comparison methods to Appendix~\ref{sec:full_real_comp}, 
as this experiment offers limited scope for meaningful comparison.

Since the missing outcomes are unobserved, we cannot directly evaluate the coverage or informativeness of the prediction intervals.
Those aspects are more directly assessed in the simulation and semi-synthetic experiments. Instead, we provide a heuristic assessment of the interval widths by comparing them with the empirical spread of the observed outcomes. Table~\ref{tab:empirical-application} reports the prediction interval widths together with the central 90\% and 95\% spans of the observed outcomes in each dataset.
To be clear, these spans do not guarantee coverage at their respective levels, since they are constructed from the observed group, which may differ from the unobserved group. Instead, they should be viewed simply as empirical measures of outcome variability.

\begin{table}[!htbp]
\centering\small
\setlength{\tabcolsep}{5pt}
\begin{tabular}{lrrrr}
\toprule
& \multicolumn{2}{c}{Observed outcome span} & \multicolumn{2}{c}{IV method} \\
\cmidrule(lr){2-3} \cmidrule(lr){4-5}
Dataset & Central 90\% & Central 95\% & Interval width & $p_{\rm inf}$ (\%) \\
\midrule
FES & 1.49 & 1.85 & 2.11 & 0.0 \\
NHANES III & 13.00 & 14.00 & 7.61 & 0.0 \\
\bottomrule
\end{tabular}
\caption{Full empirical experiment at level $1-\alpha=0.9$: empirical central 90\% and 95\% ranges among respondents and prediction interval widths from Procedure~\ref{alg:policy}, with $p_{\rm inf}$ denoting the percentage of infinite intervals.}
\label{tab:empirical-application}
\end{table}

For FES, the prediction intervals have width 2.11, about 1.42 and 1.14 times the central 90\% and 95\% spans of the observed log wages, respectively. 
Thus, the interval widths generally follow 
the scale of the observed outcomes.
We also consider the estimated outcome mass retained under the learned policy, detailed in Appendix~\ref{sec:empirical-policy-mass}. This quantity estimates the mass captured by the fitted policy, whose population counterpart is bounded above by the identification index $\Gamma_P$. Thus, an estimate above $1-\alpha$ provides heuristic evidence that informative inference is possible at the target level. For FES, this estimate is 0.910, only slightly above the target level of 0.9. Hence, although the estimate suggests that informative inference is possible, the procedure must retain nearly all of the mass captured by the learned policy to reach the target, which is consistent with the relatively wide prediction intervals.

For NHANES III, all 23 prediction intervals are finite and have width 7.61 inches, about 0.59 and 0.54 times the central 90\% and 95\% spans of the observed reported heights, respectively. Thus, the intervals are relatively narrow compared with the empirical spread of the observed outcomes. The estimated adjusted policy mass is 0.959, leaving substantially more room above the target level of 0.9 than for FES. This helps explain why the fitted procedure can attain the target coverage with narrower intervals relative to the observed outcome spread.

\section{Discussion}

In this work, we studied the problem of distribution-free predictive inference with outcomes missing not at random in the presence of instrumental variables. Our results show that the possibility of informative inference is characterized
the \emph{identification index}. 
We introduce a method, policy-based inference, which can provide meaningful inference when informative inference is possible.

Many open questions remain. One direction is to explore the broader implications of the identification index for instrumental variable problems beyond the predictive inference problem considered here. In particular, it may offer a useful perspective on characterizing and understanding the limits of identification in causal inference and other partially identified models.

In predictive inference, one important direction is to strike an appropriate balance between the strength of the assumptions and that of the inferential target. While this work focuses on the distribution-free setting, additional assumptions may lead to tighter control of the coverage. Another direction is to extend the framework to causal inference, for example, to predictive inference for the individual treatment effect of a new test individual. This extension is particularly challenging because, unlike the inference with the marginal guarantee considered in this work, inference on the individual treatment effect inherently involves an individual-conditional target. We leave these directions for future work.

\section*{Acknowledgment}
This work was supported in part by 
NIH R01-AG065276, R01-GM139926, NSF 2210662, P01-AG041710, R01-CA222147, as well as the ARO, ONR, and the Sloan Foundation.

\setlength{\bibsep}{3pt plus 0.3ex}
\bibliographystyle{plainnat-abbrev}
\bibliography{ref}

\appendix

\section{Additional details}\label{sec:method_detail}

\subsection{Generalization of identification index}
\label{sec:index_general}

Here, we generalize the definition of the identification index to a setting where $F_{z,x,1}$ might not admit a density with respect to the Lebesgue measure. For every $z\in\Z_x$ and $x\in\X$, let $Q_{z,x}(B)=\PPst{R=1,Y\in B}{X=x,Z=z}$ and define $\nu_x=\sum_{z\in\Z_x}Q_{z,x}$. Let \(h_{z,x}=\frac{\mathrm{d}Q_{z,x}}{\mathrm{d}\nu_x}\), and define
\[\gamma_P(x)=\int_{\mathcal{Y}}i_{P,x}(y)\,\mathrm{d}\nu_x(y), \text{ where } i_{P,x}(y)=\max_{z\in\Z_x}h_{z,x}(y).\] Then the \emph{identification index} of $P$ can be defined as before: \(\Gamma_P = \Ep{X \sim P_X}{\gamma_P(X)}\). 

The definition of $\gamma_P(x)$, and hence $\Gamma_P$, does not depend on the choice of the dominating measure $\nu_x$. When $\mathcal{Y}=\mathbb{R}$ and the observed-outcome distributions admit Lebesgue densities, this definition reduces to the original one. We also note that the proof of Theorem~\ref{thm:impos} extends to this generalized definition of the identification index.

\subsection{Sharpness of the identifiable miscoverage bound}\label{sec:miscoverage_sharpness}

Here we provide a theoretical discussion of the sharpness of the bound $\PP{S>t}\leq \EE{\delta(t,X)}$, for the function $\delta$ defined in~\eqref{eqn:delta_def}.
As mentioned, sharpness is understood over the class of full data distributions compatible with a fixed observed data distribution, for a fixed score $s$ and threshold $t$.

\begin{proposition}\label{prop:sharp}
Suppose that $\leb{\{y:s(x,y)>t\}}>0$ for $P_X$-almost every $x$ with $\gamma_P(x)<1$, for every finite $t$. Then
\[
\sup_{\substack{Q\in\Piv:\\Q_{X,Z,R,RY}=P_{X,Z,R,RY}}}
\Pp{Q}{s(X,Y)>t}=\Ep{P}{\delta(t,X)}.
\]
\end{proposition}
The proof is deferred to Section~\ref{sec:proofs}.
Therefore, under the condition of Proposition~\ref{prop:sharp}, there is no smaller upper bound that depends only on the observed data distribution.

\subsection{Failure of pathwise differentiability of the tight bound}\label{sec:envelope_nonregularity}

For a fixed score $s$ and threshold $t$, define
\[
\Theta_t(P)=\Ep{P}{\delta_P(t,X)},
\qquad
\delta_P(t,x)=1-\int_{\R}i_{P,x}(y)\One{s(x,y)\le t}\,\mathrm{d}y,
\]
where we use the notation $\delta_P$ with the subscript $P$ to make its dependence on $P$ explicit.  Since $\Theta_t$ depends only on the observed data distribution, pathwise differentiability is considered with respect to the model of distributions of $O=(X,Z,R,RY)$ induced by $\Piv$.

A related failure of pathwise differentiability at tied maxima arises in the context of optimal treatment values; see \citet[Theorem~1]{luedtke2016optimal}. The following result provides a direct construction for the present functional. Although the result is stated for the specific threshold $t=1/2$, similar arguments apply to other thresholds.

\begin{theorem}[Failure of pathwise differentiability]\label{thm:envelope_nonregularity}
For the score $s(x,y)=|y|$ and threshold $t=1/2$, there exists a distribution $P_0\in\Piv$ under which $\Theta_{1/2}$ is not pathwise differentiable in the observed data model, even when $X$ is constant, the instrument $Z$ is binary with known propensity $1/2$, and the response probabilities are bounded away from zero and one. Consequently, $\Theta_{1/2}$ admits no influence function at $P_0$.
\end{theorem}

\begin{proof}
Consider $\varepsilon \in (-1/4,1/4)$.
Let $P_\varepsilon$ have $X$ constant, $Z\sim\operatorname{Bernoulli}(1/2)$, and $Y\sim\operatorname{Unif}[-1,1]$ independently of $Z$, with
\[
\Pp{P_\varepsilon}{R=1\mid X,Z=z,Y=y}
=\frac12+(1-2z)\varepsilon,\qquad z\in\{0,1\}.
\]
Every $P_\varepsilon$ satisfies the IV assumption, and its response probabilities lie in $(1/4,3/4)$. Writing $f(y)=\frac12\One{|y|\le1}$ and suppressing the constant covariate, the respondent subdensities are
\[q_{z,\varepsilon}(y)=\left\{\frac12+(1-2z)\varepsilon\right\}f(y),\qquad z\in\{0,1\},\]
so that
\[i_\eps(y) := i_{P_\varepsilon,x}(y)=\max_{z\in\{0,1\}}q_{z,\varepsilon}(y)=\left(\frac12+|\varepsilon|\right)f(y).
\]
These two instrument-specific subdensities coincide at $\varepsilon=0$, and the maximizing instrument depends on the sign of $\varepsilon$. Therefore,
\[
\Theta_{1/2}(P_\varepsilon)
=1-\int_{-1/2}^{1/2}i_\varepsilon(y)\,\mathrm{d}y
=\frac34-\frac12|\varepsilon|.
\]
Let $P_\varepsilon^O$ denote the distribution of $O$ under $P_\varepsilon$. The likelihood ratio is given by
\[
\frac{\mathrm{d}P_\varepsilon^O}{\mathrm{d}P_0^O}(O)
=1+\varepsilon G(O),
\quad \text{ where }\; G(O)=2(1-2Z)(2R-1).
\]
Since $G$ is bounded, $\Ep{P_0}{G}=0$, and $\Ep{P_0}{G^2}=4$, the submodel is differentiable in quadratic mean at zero with score $G$. If $\Theta_{1/2}$ were pathwise differentiable at $P_0$, there would exist a mean-zero, square-integrable influence function $\phi(O)$ such that
\[
\Theta_{1/2}(P_\varepsilon)
=\Theta_{1/2}(P_0)+\varepsilon\Ep{P_0}{\phi(O)G(O)}+o(|\varepsilon|).
\]
However, as shown above, $\Theta_{1/2}(P_\varepsilon)=3/4-|\varepsilon|/2$, and its right and left derivatives at zero are $-1/2$ and $1/2$, respectively, contradicting the existence of the above expansion. This proves the claim.
\end{proof}

\subsection{Lemmas for policy-based inference}

The following lemma shows that the function $\tilde{\delta}_a$ defined in~\eqref{eqn:delta_policy} satisfies the assumptions of Lemma~\ref{lem:coverage_dr}.

\begin{lemma}\label{lem:policy_conditions}
The map $t\mapsto\EE{\tilde{\delta}_a(t,X)}$ is continuous and nonincreasing. Moreover, \(\EE{\tilde{\delta}_a(t,X)}\geq \EE{\delta(t,X)} \geq \PP{S > t}\) for every $t\in\R$.
\end{lemma}

Therefore, we find the efficient influence function of $\Psi_{a,t}(P) = \EE{\tilde{\delta}_a(t,X)}$ and follow the steps in Section~\ref{sec:dr_eif}. We show the following.
\begin{lemma}\label{lem:policy_eif}
Let $\psi_{a,t}(O,\eta)$ be defined as in~\eqref{eqn:policy_eif}. Then for nuisance estimators $\hat{\eta}=(\hat{\kappa}_{z,x},\hat{m}_{z,x}^a)_{z\in\Z, x \in \X}$,
\[R_{a,t}(\hat{\eta},\eta) := \EE{\psi_{a,t}(O;\hat{\eta})}-\Psi_{a,t}(P) =-\sum_{z \in \Z}\EE{\frac{\hat{\kappa}_{z,X}-\kappa_{z,X}}{\hat{\kappa}_{z,X}}\left(\hat{m}_{z,x}^a(t)-m_{z,X}^a(t)\right)}.\]
\end{lemma}

Note that the remainder term $R_{a,t}(\hat{\eta},\eta)$ is given by the product of the estimation errors of $\kappa_{z,x}$ and $m_{z,x}^a$, implying a doubly robust property. Combining the results of Lemmas~\ref{lem:policy_conditions} and~\ref{lem:policy_eif} with Lemma~\ref{lem:coverage_dr}, we obtain the coverage guarantee in Theorem~\ref{thm:policy_coverage} for the prediction set~\eqref{eqn:pred_set_policy}.

\subsection{Upper bound on coverage}\label{sec:upper_bound}

Here, we derive an upper bound on the coverage probability of the policy-based prediction set, quantifying its tightness. Let $\Gamma_P$ denote the identification index, and define the policy regret as \(\operatorname{Reg}(a) =\mathbb E\left[\int_{\mathbb R} \{i_{P,X}(y)-q_{a(X,y),X}(y)\}\,\mathrm{d}y\right]\), which quantifies the suboptimality of the chosen policy. As in the preceding analysis, we condition on the training data and treat the estimated nuisance functions and the policy as fixed. We impose the following shape restriction on the estimated nuisance functions.

\begin{assumption}\label{asm:policy_upper_shape}
For each $z\in\mathcal Z$ and almost every $x$, the function
$t\mapsto\hat m_{z,x}^a(t)$ is nondecreasing and right-continuous, with values in $[0,1]$.
\end{assumption}

We then obtain the following upper bound on the coverage probability of the prediction set from Algorithm~\ref{alg:policy}.

\begin{theorem}\label{thm:policy_coverage_upper}
Suppose $t\mapsto\Psi_{a,t}(P)$ is continuous, $b\le\hat\kappa_{z,X}\le1$ almost surely for every $z\in\Z$, and Assumption~\ref{asm:policy_upper_shape} holds. Then, for $n\ge1$,
\[\PP{Y_{n+1}\in\Ch(X_{n+1})}
\le1-\alpha+(1-\Gamma_P)+\operatorname{Reg}(a)+B_a
+\left(|\Z|+\frac2b-2\right)\sqrt{\frac{\pi}{8n}},
\]
where
\[
B_a=\frac1b\sum_{z\in\mathcal Z}\|\hat\kappa_{z,X}-\kappa_{z,X}\|_2\cdot\sup_{t\in\R}\|\hat m_{z,X}^a(t)-m_{z,X}^a(t)\|_2.
\]
\end{theorem}

The upper bound captures three sources of conservativeness: identification uncertainty, policy regret, and estimation and calibration errors. In particular, for an optimal policy with $\operatorname{Reg}(a)=0$, the excess coverage is bounded by $1-\Gamma_P$ up to a doubly robust nuisance-error term and an $O(n^{-1/2})$ calibration error. This is consistent with the empirical observation that prediction sets become less tight as $\Gamma_P$ decreases.

\subsection{Identification index of the simulation distribution}\label{sec:dgp_id}
We verify that the parameter $\Gamma$ in the distribution specified in Section~\ref{sec:dgp} coincides with the identification index $\Gamma_P$. By the definition of $\bar h$, we have \(\EEst{h_x(Y)-\bar h}{X=x,Z=z}=0\). Therefore, under the missingness mechanism specified above,
\begin{align*}
\PPst{R=1}{X=x,Z=z}
&=\EEst{\PPst{R=1}{X=x,Z=z,Y}}{X=x,Z=z}\\
&=\pi_{z,x}+\rho\frac{\pi_{z,x}}{\Gamma}(\Gamma-\pi_{z,x})\EEst{h_x(Y)-\bar h}{X=x,Z=z}=\pi_{z,x}=\Gamma w_{z,x}.
\end{align*}
Thus, writing $f_x(y) = f_{Y \mid X=x}(y)$ for simplicity,
\begin{multline*}
i_{P,x}(y)=\max_{z\in\mathcal Z}\pi_{z,x}f_{z,x,1}(y)=\max_{z\in\mathcal Z}f_x(y)\PPst{R=1}{X=x,Z=z,Y=y}\\
=\Gamma f_x(y)\max_{z\in\mathcal Z}w_{z,x}\{1+\rho(1-w_{z,x})(h_x(y)-\bar h)\}=\Gamma f_x(y),
\end{multline*}
where the last equality holds since $\max_z w_{z,x}=1$ for all $x$ by design, and $w\mapsto w\{1+\rho(1-w)(h-\bar h)\}$ is increasing on $[0,1]$ under both settings. It follows that \(\gamma_P(x)=\int_{\mathbb R}i_{P,x}(y)\,\mathrm{d}y=\int_{\mathbb R}\Gamma f_x(y)\,\mathrm{d}y=\Gamma\), which implies $\Gamma_P=\Gamma$. 

\subsection{Data generating distributions}
Below are detailed descriptions of the data-generating distributions used in the experiments in Section~\ref{sec:wcp}. In each setting, $\varepsilon$ is independent of $(X,Z)$, so the conditional outcome distribution does not depend on the instrument.

\paragraph{Setting 1}
\[
\begin{aligned}
&X\sim N_p(0,\Sigma),\qquad \Sigma_{jk}=0.5^{|j-k|},\qquad j,k=1,\ldots,p,\qquad p=10,\quad K=10,\\
&\PPst{Z=z}{X=x}=
\frac{\exp\{b_z+x_1c_z+x_2d_z\}}
{\sum_{k=0}^{K-1}\exp\{b_k+x_1c_k+x_2d_k\}},
\qquad z=0,\ldots,K-1,\\
&\mu(x)=x_1+0.8\sin(x_2)+0.5x_3^2+0.5x_1x_2,\qquad
Y=\mu(X)+\varepsilon,\qquad \varepsilon\sim N(0,1),\\
&\PPst{R=1}{Z=z,\varepsilon}=
\begin{cases}
0.995-0.005\exp(-\varepsilon^2),&z=1,\\
0.4\exp(-\varepsilon^2),&z\neq 1,
\end{cases}
\end{aligned}
\]
where \(b_z\in[0.4,-0.4]\), \(c_z\in[-1,1]\), and \(d_z\in[-0.2,0.2]\), with each sequence equally spaced over its corresponding interval.

\paragraph{Setting 2}
\[
\begin{aligned}
&X\sim N_p(0,\Sigma),\qquad \Sigma_{jk}=0.5^{|j-k|},\qquad j,k=1,\ldots,p,\qquad p=10,\quad K=10,\\
&\PPst{Z=z}{X=x}=
\frac{\exp\{b_z+x_1c_z+x_2d_z\}}
{\sum_{k=0}^{K-1}\exp\{b_k+x_1c_k+x_2d_k\}},
\qquad z=0,\ldots,K-1,\\
&\mu(x)=x_1+0.8\sin(x_2)+0.5x_3^2+0.5x_1x_2,\qquad
Y=\mu(X)+\varepsilon,\qquad \varepsilon\sim N(0,1),\\
&\PPst{R=1}{Z=z,\varepsilon}=
\begin{cases}
0.9-0.03\exp(-\varepsilon^2),&z=1,\\
0.4\exp(-\varepsilon^2),&z\neq 1.
\end{cases}
\end{aligned}
\]
where \(b_z\in[0.4,-0.4]\), \(c_z\in[-1,1]\), and \(d_z\in[-0.2,0.2]\), with each sequence equally spaced over its corresponding interval.

\paragraph{Setting 3}
\[
\begin{aligned}
&X\sim N_p(0,\Sigma),\qquad \Sigma_{jk}=0.5^{|j-k|},\qquad j,k=1,\ldots,p,\qquad p=10,\quad K=10,\\
&\PPst{Z=z}{X=x}=
\frac{\exp\{b_z+x_1c_z+x_2d_z\}}
{\sum_{k=0}^{K-1}\exp\{b_k+x_1c_k+x_2d_k\}},
\qquad z=0,\ldots,K-1,\\
&\mu(x)=x_1+0.8\sin(x_2)+0.5x_3^2+0.5x_1x_2,\qquad
Y=\mu(X)+\varepsilon,\qquad \varepsilon\sim N(0,1),\\
&\PPst{R=1}{Z=z,\varepsilon}=
\begin{cases}
0.90-0.03\exp(-\varepsilon^2),&z=1,\\
0.80+0.03\exp(-\varepsilon^2),&z\neq 1.
\end{cases}
\end{aligned}
\]
where \(b_z\in[0.4,-0.4]\), \(c_z\in[-0.1,0.1]\), and \(d_z\in[-0.02,0.02]\), with each sequence equally spaced over its corresponding interval. The identification index values are approximately 0.99, 0.88, and 0.88 for Settings 1, 2, and 3, respectively.

\subsection{Coverage guarantee for the naive plug-in prediction set}
\label{app:plug_in}

\begin{theorem}\label{thm:plug_in}
    Suppose that the function $t\mapsto \hat{\delta}(t,x)$ is nonincreasing for all $x \in \X$ and takes values in $[0,1]$. Then the prediction set $\Ch^{\mathrm{naive}}(\cdot)$ in~\eqref{eqn:pred_set_plugin} satisfies
\begin{equation*}
    \PP{Y_{n+1} \in \Ch^{\mathrm{naive}}(X_{n+1})} \geq 1-\alpha - \sup_{t \in \R} \EE{\big|\hat{\delta}(t,X) - \delta(t,X)\big|}.
    \end{equation*}
\end{theorem}

\begin{proof}
Throughout the proof, we condition on the training data used to construct $\hat{\delta}$ and hence treat $\hat{\delta}$ as fixed. Let \(\Delta=\sup_{t\in\R}\EE{\big|\hat{\delta}(t,X)-\delta(t,X)\big|}\). For finite $t$, define $\hat{\delta}^+(t,x)=\lim_{u \rightarrow t+}\hat{\delta}(u,x)$, and extend it by \(\hat{\delta}^+(-\infty,x)=\lim_{t\to-\infty}\hat{\delta}(t,x)\) and \(\hat{\delta}^+(+\infty,x)=0\). For any finite $t$, take a monotone decreasing sequence $(t_k)_{k \in \N}$ converging to $t$. Then by the right-continuity of $\delta$ and the dominated convergence theorem,
\[
\EE{\big|\hat{\delta}^+(t,X)-\delta(t,X)\big|}
=\lim_{k\to\infty}\EE{\big|\hat{\delta}(t_k,X)-\delta(t_k,X)\big|}
\le\Delta.
\]
Similarly, taking $(t_k)_{k \in \N}$ converging to $-\infty$, since $\lim_{t \rightarrow -\infty}\delta(t,x) =  1$, we have
\[
\EE{\big|\hat{\delta}^+(-\infty,X)-\delta(-\infty,X)\big|}
=\lim_{k\to\infty}\EE{\big|\hat{\delta}(t_k,X)-\delta(t_k,X)\big|}
\le\Delta.
\]
Finally, since $\hat{\delta}^+(+\infty,x)=\delta(+\infty,x)=0$, we have \(\sup_{t\in[-\infty,+\infty]} \EE{\big|\hat{\delta}^+(t,X)-\delta(t,X)\big|} \le\Delta\). Now note that replacing $\hat{\delta}$ by its right-continuous version $\hat{\delta}^+$ does not change the infimum defining either cutoff, by monotonicity. Let
\[
\tilde{t}_\alpha = \inf\left\{t\in\R:\frac{1}{n+1}\left(\sum_{i=1}^{n}\hat{\delta}(t,X_i)+\hat{\delta}(t,X_{n+1})\right)\le\alpha\right\}.
\]
Since $\tilde{t}_\alpha$ is a symmetric function of $X_1,\ldots,X_{n+1}$, which are exchangeable, we have
\[
\EE{\hat{\delta}^+(\tilde{t}_\alpha,X_{n+1})}
=\EE{\frac{1}{n+1}\left(\sum_{i=1}^{n}\hat{\delta}^+(\tilde{t}_\alpha,X_i)+\hat{\delta}^+(\tilde{t}_\alpha,X_{n+1})\right)}
\le\alpha,
\]
where the last inequality follows from the right-continuity of $\hat{\delta}^+$ and the extended-cutoff conventions.

Next, since $\hat{t}_\alpha$ is independent of $X_{n+1}$, we have
\begin{align*}
\PP{S_{n+1}>\hat{t}_\alpha}\le\EE{\delta(\hat{t}_\alpha,X_{n+1})}\le\EE{\hat{\delta}^+(\hat{t}_\alpha,X_{n+1})}+\Delta\le\EE{\hat{\delta}^+(\tilde{t}_\alpha,X_{n+1})}+\Delta\le\alpha+\Delta,
\end{align*}
where the third inequality holds since $\tilde{t}_\alpha\le\hat{t}_\alpha$ deterministically and the function $t \mapsto \hat{\delta}^+(t,x)$ is monotone. This completes the proof.
\end{proof}

\section{The boundary case \texorpdfstring{$\Gamma_P=1-\alpha$}{}}\label{sec:corner_case}

 Here, we investigate the boundary case $\Gamma_P=1-\alpha$ and show that the possibility of informative inference depends on the support of the identified subdensity $i_{P,x}$.

\begin{theorem}\label{thm:corner_case}
Suppose $P\in\mathcal P_{\mathrm{IV}}$ satisfies $\Gamma_P=1-\alpha$, and the observed-outcome subdistributions admit Lebesgue densities. Define \(A_P(x)=\{y\in\mathbb R:i_{P,x}(y)>0\}\) and \(\Lambda_P=\EE{\leb{A_P(X)}}\). Then
\[\inf_{\mathcal A\in\Omega_\alpha} \EE{\leb{\widehat C_n(X_{n+1})}} =\Lambda_P,\]
where $\Omega_\alpha$ denotes the class of oracle algorithms satisfying distribution-free coverage at level $1-\alpha$.
\end{theorem}

\begin{proof}
For the upper bound, consider the oracle prediction set $\Ch(x)=A_P(x)$. Since $f_{Y\mid X=x}(y)\ge i_{P,x}(y)$, 
\[\PP{Y\in\Ch(X)} \ge \EE{\int_{A_P(X)}i_{P,X}(y)\,\mathrm{d}y} =\Gamma_P=1-\alpha.\] 
Thus, this set achieves the desired coverage with expected Lebesgue measure $\Lambda_P$. To ensure distribution-free validity, we define the oracle algorithm to return $\R$ whenever the observed data distribution differs from that of $P$.

For the lower bound, consider any valid oracle procedure $\Ch$ with finite expected Lebesgue measure. For each $M>0$, consider a distribution $Q^M\in\Piv$ with the same observed data distribution as $P$ and conditional outcome density \(f^M_x(y)=i_{P,x}(y)+(1-\gamma_P(x))\frac{\One{|y|\le M}}{2M}\). The existence of such a distribution follows from the construction in the proof of Theorem~\ref{thm:impos}.

Since $\Ch$ has the same distribution under $P$ and $Q^M$, from its distribution-free validity, we have
\[
1-\alpha
\le \EE{\int i_{P,X}(y) \cdot \One{y \in \Ch(X)}\,\mathrm{d}y}
+\EE{\frac{1-\gamma_P(X)}{2M}\leb{\Ch(X)\cap[-M,M]}}.
\]
The second term is bounded by $\EE{\leb{\Ch(X)}}/2M$ and hence converges to zero as $M$ grows. Therefore,
\[
1-\alpha = \Gamma_P\leq\EE{\int i_{P,X}(y) \cdot \One{y \in \Ch(X)}\,\mathrm{d}y}\leq\EE{\int i_{P,X}(y)\,\mathrm{d}y} = \Gamma_P,
\]
which implies \(\EE{\int i_{P,X}(y) \cdot \One{y \notin \Ch(X)}\,\mathrm{d}y} = 0\) almost surely.
Consequently, $A_P(X)\subseteq\Ch(X)$ up to a Lebesgue-null set almost surely, and thus \(\EE{\leb{\Ch(X)}}\ge\Lambda_P\). The inequality holds trivially for procedures with infinite expected Lebesgue measure, and thus this completes the proof.
\end{proof}

\section{Exact finite-sample coverage with known instrument propensity}\label{sec:wcp_known_pzx}

In this section, we briefly discuss an alternative procedure based on weighted conformal prediction (WCP)~\citep{tibshirani2019conformal}, for the setting where the instrument propensity $\kappa_{z,x}=\PPst{Z=z}{X=x}$ is known.

Suppose $\kappa_{\min}:=\inf_{z,x}\kappa_{z,x}>0$ holds. Let $a:\mathcal X\times\mathbb R\to\mathcal Z$ be a policy and $s:\mathcal X\times\mathbb R\to\mathbb R$ be a nonconformity score, both constructed using independent training data. Define \(W=R\cdot\One{Z=a(X,Y)} / \kappa_{Z,X}\), which can be computed from the observation $(X,Z,R,RY)$ by setting $W=0$ when $R=0$. Under Assumption~\ref{asm:iv}, we have
\[
\EEst{W}{X=x,Y=y}=\PPst{R=1}{X=x,Y=y,Z=a(x,y)}=\frac{q_{a(x,y),x}(y)}{f_{Y\mid X=x}(y)}.
\]
Consequently, the probability mass captured by the policy $a$ is
\[
\theta_a:=\EE{W}=\EE{\int q_{a(X,y),X}(y)\,dy}\le \EE{\int \max_{z\in\mathcal Z}q_{z,X}(y)\,dy}=\Gamma_P.
\]

Given calibration data independent of the training data, consider the samples $\mathcal M=\{i:R_i=1,\ Z_i=a(X_i,Y_i)\}$ whose instrument $Z_i$ matches the value of the policy $a(X_i,Y_i)$.
By Assumption~\ref{asm:iv}, the probability of this matching event conditional on $(X,Y)=(x,y)$ is
\begin{multline*}
    \PPst{R=1,\ Z=a(X,Y)}{X=x,Y=y}=\kappa_{a(x,y),x}\cdot\PPst{R=1}{X=x,Y=y,Z=a(x,y)}\\
    =\kappa_{a(x,y),x}\cdot\frac{q_{a(x,y),x}(y)}{f_{Y\mid X=x}(y)}.
\end{multline*}
Therefore, conditional on $|\mathcal M|$, the observations $(X_i,Y_i)_{i\in\mathcal M}$ are i.i.d. from a distribution with density proportional to $f_X(x)\cdot\kappa_{a(x,y),x}\cdot q_{a(x,y),x}(y)$---where $f_X$ denotes the density of $X$.

To apply WCP, when $\theta_a>0$, we introduce a policy-induced target distribution with density $f_X(x)q_{a(x,y),x}(y)/\theta_a$, which removes the effect of the instrument propensity from the policy-matched calibration distribution. Then the density ratio of this target distribution to the policy-matched calibration distribution is proportional to $w(x,y)=1/\kappa_{a(x,y),x}$.

Thus, WCP with weights $w(X_i,Y_i)$ for $i\in\mathcal M$ and candidate weight $w(x,y)$ yields a prediction set $\widehat C_\beta(x)$ satisfying
\[
\Pp{(X_i,Z_i,R_i,Y_i)_{i\in[n]} \iidsim P_{X,Z,R,Y},\,
(X_{n+1},Y_{n+1})\sim\overline Q_a}
{Y_{n+1}\in\widehat C_\beta(X_{n+1})}
\ge 1-\beta,
\]
where $\overline Q_a$ denotes the probability measure with density $f_X(x)q_{a(x,y),x}(y)/\theta_a$. Since $q_{a(x,y),x}(y)\leq f_{Y\mid X=x}(y)$, the density of $\overline Q_a$ is bounded above by $f_{X,Y}(x,y)/\theta_a$. Consequently,
\[
\Pp{(X_i,Z_i,R_i,Y_i)_{i\in[n+1]} \iidsim P_{X,Z,R,Y}}{Y_{n+1}\in\widehat C_\beta(X_{n+1})}
\ge\theta_a(1-\beta).
\]
If $\theta_a>1-\alpha$ were known, choosing $\beta=1-(1-\alpha)/\theta_a$ would therefore guarantee coverage of at least $1-\alpha$.

In practice, $\theta_a$ can be replaced by a lower confidence bound. Using an independent sample $\mathcal D_1$ of size $n_1$---which can be constructed, e.g., through additional data splitting---define
\[
\widehat\theta_L
=
\frac{1}{n_1}\sum_{i\in\mathcal D_1}
\frac{R_i\One{Z_i=a(X_i,Y_i)}}{\kappa_{Z_i,X_i}}
-
\frac{1}{\kappa_{\min}}
\sqrt{\frac{\log(1/\varepsilon)}{2n_1}},
\]
for a fixed $\varepsilon\in(0,\alpha)$. By Hoeffding's inequality, $\PP{\widehat\theta_L\le\theta_a}\ge1-\varepsilon$. If $\widehat\theta_L\le(1-\alpha)/(1-\varepsilon)$, we return $\mathbb R$. Otherwise, we apply WCP with \(\widehat\beta=1-\frac{1-\alpha}{(1-\varepsilon)\widehat\theta_L}\). For each $z\in\mathcal Z$, define \(\tau_z(x)=\operatorname{Quantile}_{1-\widehat\beta}\left(\sum_{i\in\mathcal M}p_i^{z,x}\delta_{S_i}+p_{n+1}^{z,x}\delta_{+\infty}\right)\), where $S_i=s(X_i,Y_i)$ and
\[
p_i^{z,x}=\frac{1/\kappa_{Z_i,X_i}}{\sum_{j\in\mathcal M}1/\kappa_{Z_j,X_j}+1/\kappa_{z,x}},\qquad
p_{n+1}^{z,x}=\frac{1/\kappa_{z,x}}{\sum_{j\in\mathcal M}1/\kappa_{Z_j,X_j}+1/\kappa_{z,x}}.
\]
Then the prediction set is constructed as
\begin{equation}\label{eqn:pred_set_wcp_known}
\ch(x)=
\begin{cases}
\mathbb R, & \widehat\theta_L\le(1-\alpha)/(1-\varepsilon),\\
\{y\in\mathbb R:s(x,y)\le\tau_{a(x,y)}(x)\}, & \widehat\theta_L>(1-\alpha)/(1-\varepsilon).
\end{cases}
\end{equation}
When $\mathcal M$ is empty, the weighted quantile is $+\infty$, so the prediction set is also $\mathbb R$.

\begin{theorem}
Suppose Assumption~\ref{asm:iv} holds, $\kappa_{z,x}$ is known with $\kappa_{\min}>0$, and the training, calibration, and lower-bound estimation samples are independent. Then the prediction set $\ch(X_{n+1})$ in~\eqref{eqn:pred_set_wcp_known} satisfies \(\PP{Y_{n+1}\in\ch(X_{n+1})}\ge1-\alpha\).
\end{theorem}

\begin{proof}
Let $G=\{\widehat\theta_L\le\theta_a\}$. On $G$, conditional on the training and lower-bound estimation samples, the prediction set has coverage one if $\widehat\theta_L\le(1-\alpha)/(1-\varepsilon)$. Otherwise, the WCP guarantee gives \(\PP{Y_{n+1}\in\ch(X_{n+1})} \ge\theta_a(1-\widehat\beta) \ge(1-\alpha)/(1-\varepsilon)\), where the probability is taken over the calibration and test observations. Since $\PP{G}\ge1-\varepsilon$, the unconditional coverage is at least $1-\alpha$.
\end{proof}

Although this approach achieves exact finite-sample coverage, it may be conservative in practice due to the lower confidence bound for $\theta_a$, the restriction to policy-matched respondents, and the additional sample splitting. We present this approach primarily to illustrate the possibility of exact finite-sample coverage control, rather than as a practical alternative to our proposed method.

\subsection{Simulation results}

We consider two data-generating settings with $\Gamma_P=0.95$. In both settings, we generate $R\mid X=x,Z=z,Y=y\sim\mathrm{Bernoulli}(r_z(x,y))$, where
\[
r_z(x,y)=\pi_{z,x}+\rho\frac{\pi_{z,x}}{\Gamma}(\Gamma-\pi_{z,x})\{h_x(y)-\bar h\},\qquad
\pi_{z,x}=\Gamma\frac{e_{z,x}}{\max_{k\in\mathcal Z}e_{k,x}},
\]
with $h_x(y)=2\One{y>\mu(x)}-1$ and $\bar h=\EEst{h_X(Y)}{X=x}$. In both settings, $\max_z r_z(x,y)=\Gamma$, so that the identification index satisfies $\Gamma_P=\Gamma$. We consider the following two data distributions. In Setting 1, the instrument $Z$ has $K=10$ categories, whereas Setting 2 considers a binary instrument.

\paragraph{Setting 1.} We generate $X\sim \mathcal{N}(0,I_{10})$ and $Z\in\{1,\ldots,10\}$ with \(e_{z,x}=0.05+0.5\frac{b_z\exp(\theta_z^\top x)}{\sum_{k=1}^{10}b_k\exp(\theta_k^\top x)}\), where $b_1=0.4$, $b_z=0.6/9$ for $z=2,\ldots,10$, $\theta_1=(0.16,-0.11,0.08,0,\ldots,0)^\top$, and $\theta_z=\frac{z-1}{9}(-0.05,0.04,-0.03,0,\ldots,0)^\top$ for $z=2,\ldots,10$. The outcome follows 
\[Y\mid X=x\sim 0.9\cdot\mathcal{N}(\mu(x)-1,0.25^2)+0.1\cdot\mathcal{N}(\mu(x)+9,0.25^2), \text{ where } \mu(x)=\beta^\top x,\]
with $\beta=(1,0.8,0.6,0.4,0.2,-0.2,-0.4,-0.6,-0.8,-1)^\top$. We set $\rho=0.2$ and $\bar h=-0.8$. Here, the instrument propensity satisfies $e_{z,x}\ge0.05$ for all $z,x$.

\paragraph{Setting 2.} We generate $X\sim \mathcal{N}(0,I_2)$ and $Z\in\{0,1\}$ with \(e_{1,x}=0.5+0.1\tanh(x_1)\) and \(e_{0,x}=1-e_{1,x}\). The outcome follows 
\[Y\mid X=x\sim \mathcal{N}(\mu(x),1), \text{ where } \mu(x)=x_1+0.5x_2.\]
We set $\rho=0.3$ and $\bar h=0$. The instrument propensity satisfies $e_{z,x}\ge0.4$ for all $z,x$.\\

\begin{figure}
    \centering
    \includegraphics[width=0.75\linewidth]{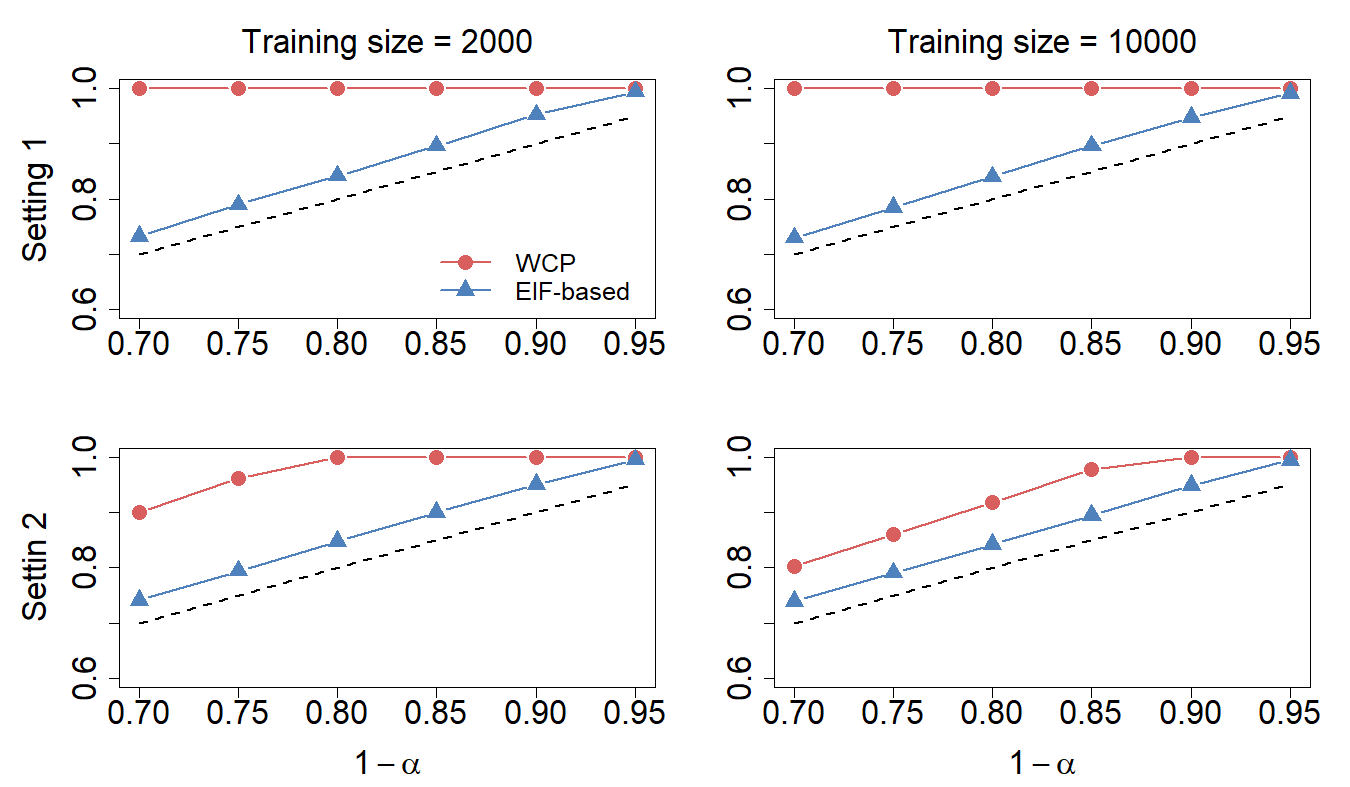}
    \caption{Coverage rates of the procedures~\eqref{eqn:pred_set_wcp_known} (WCP) and~\eqref{eqn:pred_set_policy} (EIF-based) in Settings 1 and 2, with training sizes of 2000 and 10000.}
    \label{fig:known_pzx}
\end{figure}

For each setting, we generate a training dataset of size $2000$ or $10000$, and repeat 200 trials of generating calibration and test datasets, both of size 500, applying the WCP-based procedure~\eqref{eqn:pred_set_wcp_known} and the EIF-based procedure~\eqref{eqn:pred_set_policy}, with the policy learned as described in Section~\ref{sec:policy_choice}. In this experiment, both methods leverage the information of true instrument propensity $\kappa_{z,x}$. The results are shown in Figure~\ref{fig:known_pzx}, illustrating the practical limitations of the procedure~\eqref{eqn:pred_set_wcp_known} discussed in the previous section: it fails to provide informative inference when $K$ is large and is generally conservative compared to the procedure~\eqref{eqn:pred_set_policy}.

\section{Additional simulations: Experiment with invalid IV}

Here, we repeat the experiment in Section~\ref{sec:sim_cov} with a slightly different data-generating distribution in which the instrumental variable is not valid---i.e., the condition $Z \indep Y \mid X$ does not hold---and the procedure is therefore no longer guaranteed to satisfy the coverage guarantee. We follow the overall setup of Setting 2 in Section~\ref{sec:sim_cov}, but use the following model for the outcome: \(Y = Y_0 + q \cdot \frac{2Z-11}{9}\), where $Y_0$ is the outcome generated according to the original model. Thus, the new outcome $Y$ also depends on $Z$.

\begin{figure}[htbp]
    \centering
    \includegraphics[width=0.95\linewidth]{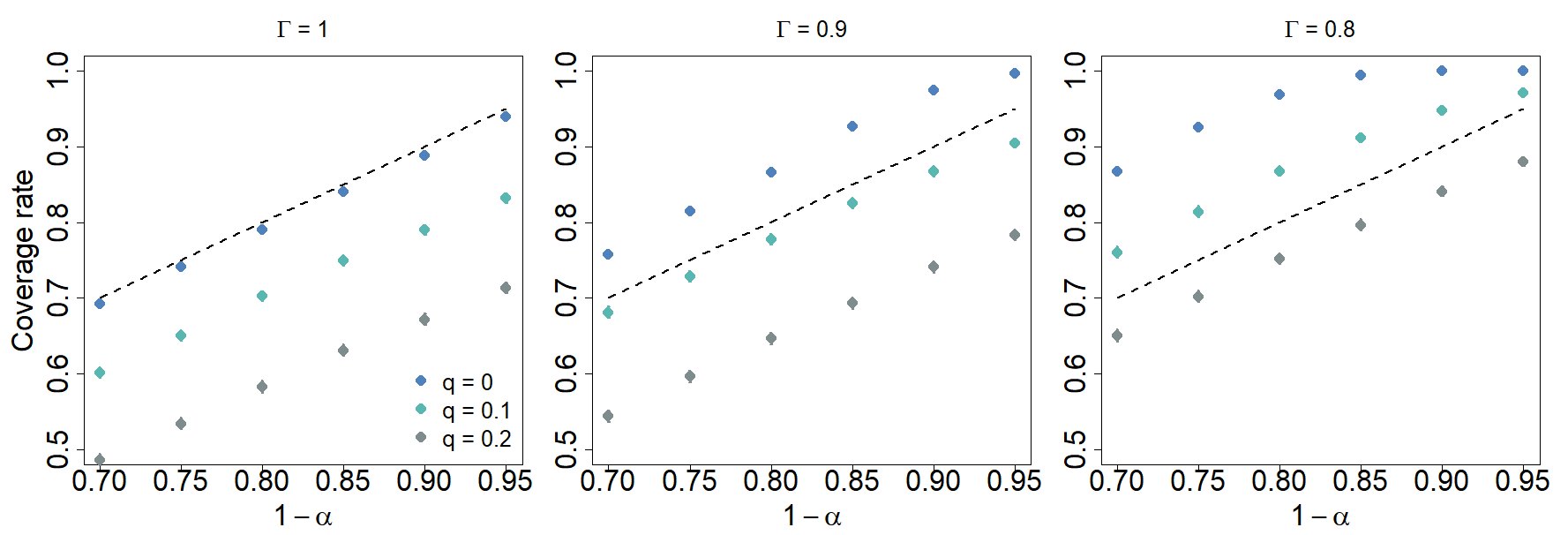}
    \caption{Coverage rates of the prediction set from Procedure~\ref{alg:policy} with true nuisance functions, under invalid instrumental variables.}
    \label{fig:invalid}
\end{figure}

We conduct the experiment with three values of the outcome-instrument dependence parameter, $q = 0, 0.1, 0.2$. Note that $q=0$ corresponds to the original valid IV setting, and that the parameter $\Gamma$ no longer represents the identification index, which is not well-defined for distributions that do not satisfy the IV assumption. We report the results for the (EIF-)oracle procedure---Procedure~\ref{alg:policy} with the true nuisance functions---to focus on the inference procedure itself and separate it from the effect of nuisance estimation quality. The results are shown in Figure~\ref{fig:invalid} and illustrate that the prediction set can undercover when the IV is not valid.

\section{Additional results for the empirical experiment}
\label{sec:empirical-application-details}

\subsection{Semi-synthetic experiment in the MAR setting}\label{sec:semi_synthetic_mar}

Here, we additionally examine the MAR setting where the instrument provides no additional information about the response. For this experiment, we generate an artificial instrument $Z$ independently and uniformly across three categories for FES and five categories for NHANES III.  We generate the missingness based on the model \(P(R=1\mid X=x,Y=y,Z=z) =\operatorname{expit}\{4+1.2u(x)\}\), where $u(x)$ is the standardized and clipped prediction of the outcome $Y$ from the estimator $\hat{\mu}(x)$, with standardization based on the mean and standard deviation of the pilot predictions and clipping to $[-3,3]$. The results are shown in Table~\ref{tab:empirical-mar}.

\begin{table}[!htbp]
\centering\small
\setlength{\tabcolsep}{3pt}
\begin{tabular}{llrrrr}
\toprule
Dataset & Method & Coverage & Mean width & Median width & $p_{\rm inf}$ (\%) \\
\midrule
FES & IV
& \cellcolor{green!20}0.928 (0.004)
& $\infty$ (--)
& 1.59 (0.044)
& 14.0 \\
 & $X$-WCP
& \cellcolor{green!20}0.902 (0.001)
& 1.40 (0.005)
& 1.39 (0.006)
& 0.0 \\
 & $(X,Z)$-WCP
& \cellcolor{green!20}0.902 (0.001)
& 1.40 (0.004)
& 1.40 (0.004)
& 0.0 \\
\addlinespace
{NHANES III} & {IV}
& \cellcolor{green!20}{0.931 (0.004)}
& {$\infty$ (--)}
& {10.41 (0.416)}
& {16.0} \\
 & {$X$-WCP}
& \cellcolor{green!20}{0.904 (0.001)}
& {9.09 (0.021)}
& {9.12 (0.068)}
& {0.0} \\
 & {$(X,Z)$-WCP}
& \cellcolor{green!20}{0.905 (0.001)}
& {9.07 (0.023)}
& {9.04 (0.017)}
& {0.0} \\
\addlinespace
\bottomrule
\end{tabular}
\caption{Semi-synthetic experiments with the FES and NHANES datasets: coverage rate, mean and median prediction interval widths, and the percentage of intervals with infinite width ($p_{\text{inf}}$) for Procedure~\ref{alg:policy} (IV), $X$-WCP, and $(X,Z)$-WCP, with standard errors, at level $\alpha=0.1$, under the MAR setting.}
\label{tab:empirical-mar}
\end{table}

As expected, the weighted conformal prediction methods achieve coverage rates close to the nominal level $1-\alpha=0.9$. Since the data satisfy the MAR assumption in this setting, WCP is expected to provide tight prediction intervals. In contrast, the policy-based inference procedure, which exploits the IV without relying on the MAR assumption, can be less informative and produce infinite-width intervals in some trials. Thus, when there is sufficient reason to believe that the outcomes are missing at random, it is natural to use WCP.

\subsection{Full empirical experiment with comparison methods}\label{sec:full_real_comp}

Table~\ref{tab:empirical-application-comparisons} shows the results of the weighted conformal prediction methods for the experiment in Section~\ref{sec:empirical-application}. Recall that WCP requires the MAR assumption and therefore does not provide a coverage guarantee under the MNAR setting considered here. Also, as discussed in Section~\ref{sec:rmk}, the proposed and comparison methods share a similar width-coverage tradeoff, and narrower intervals come at the cost of lower coverage. Thus, without observing the missing outcomes, interval width alone does not provide a meaningful comparison between the methods in the full empirical experiment.

We include two additional baselines based on best-case and worst-case completions. Starting from an oracle split conformal procedure that includes the missing calibration scores, we replace these scores by zero or $+\infty$, respectively. The resulting intervals bound the conformal prediction set that would be obtained in the hypothetical scenario where all calibration outcomes were observed. The best-case completion is optimistic and provides no coverage guarantee, whereas the worst-case completion provides a conservative population benchmark under exchangeability. At $\alpha=0.1$, the latter returns the entire real line for FES but a finite interval for NHANES III, due to its lower proportion of missing outcomes.

\begin{table}[!htbp]
\centering\small

\setlength{\tabcolsep}{5pt}
\begin{tabular}{lrrrrrr}
\toprule
& \multicolumn{3}{c}{FES} & \multicolumn{3}{c}{NHANES III} \\
\cmidrule(lr){2-4}\cmidrule(lr){5-7}
Method & Mean & Median & $p_{\rm inf}$ & Mean & Median & $p_{\rm inf}$ \\
\midrule
$X$-WCP & 1.33 & 1.33 & 0.0 & 8.11 & 8.11 & 0.0 \\
$(X,Z)$-WCP & 1.24 & 1.24 & 0.0 & 8.14 & 8.14 & 0.0 \\
Best case & 1.11 & 1.11 & 0.0 & 8.09 & 8.09 & 0.0 \\
Worst case & $\infty$ & $\infty$ & 100.0 & 9.85 & 9.85 & 0.0 \\
\bottomrule
\end{tabular}
\caption{Comparison methods for the full empirical experiment at level $1-\alpha=0.9$: mean and median interval widths and the percentage of infinite intervals ($p_{\rm inf}$) for $X$-WCP and $(X,Z)$-WCP, together with the best- and worst-case completion baselines.}
\label{tab:empirical-application-comparisons}
\end{table}

\subsection{Interpreting prediction interval widths through the learned policy}\label{sec:empirical-policy-mass}

To help interpret the widths of the prediction intervals in the full empirical experiment, we consider the amount of outcome probability mass retained under the learned policy. This quantity reflects how much of the outcome distribution can be captured by selecting an instrument value according to the fitted policy. If the retained mass is only slightly above the target level $1-\alpha$, the procedure must include nearly all of this mass to reach the target, which can result in wider prediction intervals.
Using the nuisance functions and policy fitted on the training data, we estimate this mass on the calibration data by
\begin{equation}\label{eq:empirical-policy-mass}
\begin{aligned}
\hat\theta_{\hat a}
&=\Pi_{[0,1]}\!\left\{\frac1n\sum_{i=1}^n\left[
\sum_z\widehat M_z^{\hat a}(X_i)
+\frac{R_i\One{\hat a(X_i,Y_i)=Z_i}
-\widehat M_{Z_i}^{\hat a}(X_i)}{\hat \kappa_{Z_i,X_i}}
\right]\right\}\\
&=1-\Pi_{[0,1]}\!\left(\frac1n\sum_{i=1}^n\psi_{\hat a,\infty}(O_i;\hat\eta)\right),
\end{aligned}
\end{equation}
where $\Pi_{[0,1]}(u)=\min\{1,\max\{0,u\}\}$ and \(\widehat M_z^{\hat a}(x) =\int_{\mathbb R}\One{\hat a(x,y)=z}\hat q_{z,x}(y)\,\mathrm{d}y\). The projection is applied only to the final average. The quantity $1-\hat\theta_{\hat a}$ can be interpreted as the corresponding limiting estimated miscoverage bound.

For the fits used in Section~\ref{sec:empirical-application}, we obtain $\hat\theta_{\hat a}=0.910$ for FES and $0.959$ for NHANES III, corresponding to limiting estimated miscoverage bounds of $0.090$ and $0.041$, respectively. At the target level $1-\alpha=0.9$, the FES estimate is only slightly above the target, so the procedure must retain nearly all of the estimated mass captured by the policy. In contrast, the NHANES III estimate leaves more room above the target. These estimates therefore provide some intuition for the relatively wider intervals for FES and narrower intervals for NHANES III observed in Section~\ref{sec:empirical-application}.

These quantities should only be interpreted as heuristic diagnostics. For any fixed policy $a$, the population mass captured by that policy is bounded above by $\Gamma_P$. Thus, $\hat\theta_{\hat a}$ estimates a policy-specific lower bound on the maximal attainable mass, rather than $\Gamma_P$ itself. In particular, a value of $\hat\theta_{\hat a}$ above $1-\alpha$ provides evidence that the learned policy captures sufficient mass to support informative inference at the target level, and consequently suggests that $\Gamma_P$ is also likely above $1-\alpha$. The amount by which $\hat\theta_{\hat a}$ exceeds $1-\alpha$ can further provide intuition about the informativeness of the resulting prediction intervals. However, we note that since $\hat\theta_{\hat a}$ is also estimated from the data, these comparisons are heuristic and do not constitute a formal lower bound on $\Gamma_P$ or certify the population coverage of the fitted prediction intervals.

\section{Predictive inference under a lower bound on observation probability}\label{sec:bound_obs_prob}

As an implication of our results, we consider predictive inference without an IV under the model \(\mathcal P_0=\left\{P_{X,R,Y}:\PP{R=1} > 1-\alpha \right\}\), and we observe data $(X_i,R_i,R_iY_i)_{i \in [n]}$. To see how our results apply to this setting, consider an artificial instrument $Z$ independent of $(X,R,Y)$. Then Assumption~\ref{asm:iv} clearly holds and the observed outcome subdensity satisfies $q_{z,x}(y)=q_x(y)$ for all $z\in\mathcal Z$, where $q_x(y)=f_{Y\mid X=x}(y)\cdot\PPst{R=1}{X=x,Y=y}$. Consequently, the corresponding identification index is
\[
\Gamma_P=\EE{\int\max_{z\in\mathcal Z}q_{z,X}(y)\,\mathrm{d}y}
=\EE{\int q_X(y)\,\mathrm{d}y} = \PP{R=1} > 1-\alpha.
\]
Thus, our framework implies that informative predictive inference is possible under this model, even without an IV. This is intuitive, as it states that predictive inference becomes possible when the missingness probability is small.

Motivated by this observation, we can develop an analogous oracle procedure directly under the model $\mathcal P_0$, without introducing an artificial instrument. Observe that for any nonconformity score $s(x,y)$, we have \(\PP{s(X,Y)\le t}\ge\EE{R\cdot\One{s(X,Y)\le t}}\). Thus, defining the oracle threshold
\[
t_\alpha=\inf\left\{t\in\R:\EE{R\cdot\One{s(X,Y)\le t}}\ge 1-\alpha\right\},
\]
the oracle prediction set $\Ch^{\mathrm{or}}(x)=\{y:s(x,y)\le t_\alpha\}$ achieves marginal coverage of at least $1-\alpha$---this is the specialization of~\eqref{eqn:pred_set_oracle_2} to the present model.
Moreover, since
$\lim_{t \rightarrow \infty} \EE{R\cdot\One{s(X,Y)\le t}} = \PP{R=1}>1-\alpha$, the threshold $t_\alpha$ is finite---which yields a finite-width prediction interval if, e.g., we use the standard residual score $s(x,y) = |y - \hat{\mu}(x)|$. 

We can also have a non-oracle conformal-type prediction set $\ch(x)=\{y:s(x,y)\le \hat{t}_\alpha\}$ with threshold
\[\hat{t}_\alpha=\inf\left\{t\in\mathbb R:\frac{\sum_{i=1}^n (1-R_i\cdot\One{s(X_i,Y_i)\le t}) + 1}{n+1}\le\alpha\right\},\]
with $\inf\varnothing=+\infty$, analogously to the semi-oracle prediction set $\Ch^{\mathrm{sor}}$ in~\eqref{eqn:pred_set_oracle}.

Observing a valid IV may further improve the informativeness of predictive inference, particularly when the IV is relevant, since the identification index generally satisfies \(\Gamma_P=\EE{\int\max_{z\in\mathcal Z}q_{z,X}(y)\,\mathrm{d}y}\ge\PP{R=1}\), where the inequality is strict if $Z\not\indep R\mid X$, provided that $\PPst{Z=z}{X}>0$ almost surely for every $z\in\mathcal Z$. Thus, while an IV is not necessary for informative predictive inference under $\mathcal P_0$, it may provide additional information that allows for more informative prediction sets.

\section{Proofs of theorems}\label{sec:proofs}

\paragraph{Notation}

For a distribution $F$ on $\X$ and a measurable set $A \subset \X$, we write $F(A)$ to denote the probability $\Pp{X \sim F}{X \in A}$.

\subsection{Proof of Theorem~\ref{thm:impos}}

Fix any $P \in \Piv$ satisfying $\Gamma_P < 1-\alpha$. We now construct another distribution $Q^M \in \Piv$ for an arbitrary $M \in \R$ as follows.
First observe that for any $z \in \Z$,
\[
\gamma_P(x) = \int i_{P,x}(y)\,\mathrm{d}y \geq \int \pi_{z,x} f_{z,x,1}(y)\,\mathrm{d}y = \pi_{z,x}.
\]
For each $z \in \Z$, consider two cases. If $\gamma_P(x)>\pi_{z,x}$, define \(y\mapsto g_{z,x}(y) = \frac{i_{P,x}(y) - \pi_{z,x} f_{z,x,1}(y)}{\gamma_P(x)-\pi_{z,x}}\), which is a density function by construction, and let $G_{z,x}$ be the distribution with density $g_{z,x}$. 
If $\gamma_P(x)=\pi_{z,x}$, then 
$i_{P,x}(y)=\pi_{z,x}f_{z,x,1}(y)$ almost everywhere, and we let $G_{z,x}$ be any probability distribution on $\R$.
For $\pi_{z,x}<1$, define
\[H_{z,x} = \frac{\gamma_P(x)-\pi_{z,x}}{1-\pi_{z,x}}\cdot G_{z,x} + \frac{1-\gamma_P(x)}{1-\pi_{z,x}}\cdot \delta_M.\]
If $\pi_{z,x}=1$, then $\gamma_P(x)=1$, and we let $H_{z,x}$ be any probability distribution on $\R$.

Next, for each $x \in \X$ and $z \in \Z$, let $V_{z,x}$ be a random variable on $\R$ with distribution $H_{z,x}$, where the collection $\{V_{z,x}\}_{z\in\Z,\;x\in\X}$ is independent of $(X,Z,R,Y)$. Let \(Y' = RY + (1-R)V_{Z,X}\). Note that, for the distribution of $Y'$ to be well-defined, it is not necessary to specify the conditional distribution of $V_{z,x}$ given $R=1$, since $(1-R)V_{Z,X}=0$ whenever $R=1$. We denote by $Q^M$ the distribution of $(X,Z,R,Y')$.

We now explore the properties of $Q^M$. First, observe that, by construction, 
\[RY' = R(RY + (1-R)V_{Z,X}) = R^2Y = RY\]
holds deterministically. This immediately implies that, under $P$, \((X,Z,R,RY) \stackrel{d}{=} (X,Z,R,RY')\), which essentially means that \(P_{X,Z,R,RY} = Q^M_{X,Z,R,RY}\).

Next, observe that for any $x \in \X$, $z \in \Z$, and measurable set $B \subset \R$,
\begin{align*}
&\PPst{Y' \in B}{X = x,Z = z}\\
&= \PPst{Y \in B}{R = 1, X = x,Z = z}\PPst{R = 1}{X = x,Z = z}\\
&\hspace{6cm}+ \PPst{V_{z,x} \in B}{R = 0, X = x,Z = z}\PPst{R = 0}{X = x,Z = z}\\
&= F_{z,x,1}(B) \cdot \pi_{z,x} + H_{z,x}(B)\cdot (1 - \pi_{z,x})\\
&= F_{z,x,1}(B) \cdot \pi_{z,x} + (\gamma_P(x)-\pi_{z,x})\cdot G_{z,x}(B) + (1-\gamma_P(x))\cdot \delta_M(B)\\
&= \int_B i_{P,x}(y)\,\mathrm{d}y + (1-\gamma_P(x))\cdot \delta_M(B).
\end{align*}
The last expression does not depend on $z$, implying that $Y' \indep Z \mid X$. Hence, $Q^M \in \Piv$.

In summary, we have constructed a distribution $Q^M \in \Piv$ whose distribution of the observed data coincides with that of $P$. This implies that the coverage probability of $\Ch(X_{n+1})$ for $Y_{n+1}'$---which is defined similarly to $Y'$ above---under $P$ coincides with that for $Y_{n+1}$ under $Q^M$, i.e.,
\begin{equation}\label{eqn:cov_PQ}
\Pp{(X_i,Z_i,R_i,Y_i)_{i\in[n+1]}\iidsim P}
{Y_{n+1}'\in\Ch(X_{n+1})}
=
\Pp{(X_i,Z_i,R_i,Y_i)_{i\in[n+1]}\iidsim Q^M}
{Y_{n+1}\in\Ch(X_{n+1})}.
\end{equation}
Indeed, 
$\Ch(X_{n+1})$ is a function only of the observed data and hence has the same distribution under $P$ and $Q^M$, while the conditional distribution of $Y_{n+1}'$ given $X_{n+1}$ under $P$ coincides with that of $Y_{n+1}$ given $X_{n+1}$ under $Q^M$ by the construction above.

Now, since $Q^M \in \Piv$ and $\A$ satisfies distribution-free coverage under IV, the right-hand side of~\eqref{eqn:cov_PQ} is at least $1-\alpha$. Therefore, under $P$, we have
\begin{align*}
1 - \alpha \leq \PP{Y_{n+1}'\in\Ch(X_{n+1})} = \EE{\PPst{Y_{n+1}'\in\Ch(X_{n+1})}{X_{n+1},Z_{n+1}}},
\end{align*}
For each $z \in \Z$ and $x \in \X$ such that $\pi_{z,x}<1$, let $W_{z,x}$ be a random variable with distribution \(W_{z,x} \sim \mathrm{Bernoulli}\!\left(\frac{\gamma_P(x)-\pi_{z,x}}{1-\pi_{z,x}}\right)\), and if $\pi_{z,x}=1$, let $W_{z,x}=1$. Let the collection $\{W_{z,x}\}_{z\in\Z,\;x\in\X}$ be independent of all the datapoints. Also, let $L_{z,x}$ be a random variable for each $z \in \Z$ and $x \in \X$ with distribution $G_{z,x}$, where the collection $\{L_{z,x}\}_{z\in\Z,\;x\in\X}$ is independent of all the datapoints and of $\{W_{z,x}\}_{z\in\Z,\;x\in\X}$. For $\pi_{z,x}<1$, the random variable $V_{z,x}$ can then be represented as \(V_{z,x} = W_{z,x}L_{z,x} + (1-W_{z,x})M\). When $\pi_{z,x}=1$, we have $\gamma_P(x)=1$ and $R=1$ almost surely conditional on $X=x,Z=z$, and thus the value of $V_{z,x}$ is irrelevant.

By construction, the event $E = \{R_{n+1} = 0 \text{ and } W_{Z_{n+1},X_{n+1}} = 0\}$ implies the event $\{Y_{n+1}'=M\}$. Therefore,
\begin{align*}
&\PPst{Y_{n+1}'\in\Ch(X_{n+1})}{X_{n+1},Z_{n+1}}\\
&= \PPst{Y_{n+1}'\in\Ch(X_{n+1}), E^c}{X_{n+1},Z_{n+1}} + \PPst{M\in\Ch(X_{n+1}), E}{X_{n+1},Z_{n+1}}\\
&\leq \PPst{E^c}{X_{n+1},Z_{n+1}} + \PPst{M\in\Ch(X_{n+1})}{X_{n+1},Z_{n+1}}\\
&= 1 - \PPst{E}{X_{n+1},Z_{n+1}} + \PPst{M\in\Ch(X_{n+1})}{X_{n+1},Z_{n+1}}\\
&= 1 - \PPst{R_{n+1}=0}{X_{n+1},Z_{n+1}}\PPst{W_{Z_{n+1},X_{n+1}}=0}{X_{n+1},Z_{n+1}} + \PPst{M\in\Ch(X_{n+1})}{X_{n+1},Z_{n+1}}\\
&= \gamma_P(X_{n+1}) + \PPst{M\in\Ch(X_{n+1})}{X_{n+1},Z_{n+1}},
\end{align*}
where the last equality follows directly when $\pi_{Z_{n+1},X_{n+1}}<1$. When $\pi_{Z_{n+1},X_{n+1}}=1$, we have $\gamma_P(X_{n+1})=1$ and $R_{n+1}=1$ almost surely conditional on $(X_{n+1},Z_{n+1})$, so the same equality holds regardless of the value of $W_{Z_{n+1},X_{n+1}}$.

Marginalizing both sides, we have
\[1 - \alpha \leq \PP{Y_{n+1}'\in\Ch(X_{n+1})} \leq \EE{\gamma_P(X_{n+1})} + \PP{M\in\Ch(X_{n+1})},\]
which implies \(\PP{M\in\Ch(X_{n+1})} \geq 1-\Gamma_P-\alpha\). Note that $1-\Gamma_P-\alpha > 0$ by assumption. Therefore,
\begin{multline*}
\EE{\leb{\Ch(X_{n+1})}} = \EE{\int_\R \One{M \in \Ch(X_{n+1})} \mathrm{d} M} = \int_\R \EE{\One{M \in \Ch(X_{n+1})}} \mathrm{d} M\\
= \int_\R \PP{M \in \Ch(X_{n+1})} \mathrm{d} M \geq \int_\R (1-\Gamma_P-\alpha)\, \mathrm{d} M = \infty,
\end{multline*}
where the second equality follows from Tonelli's theorem.

\subsection{Proof of Theorem~\ref{thm:coverage}}
We first prove the coverage guarantee for the oracle prediction set $\Ch^{\mathrm{or}}$ in~\eqref{eqn:pred_set_oracle_2}. For any $x$ and $t$, since $i_{P,x}(y)\le f_x(y) := f_{Y \mid X=x}(y)$, we have
\[
\PPst{s(X,Y)\le t}{X}=\int f_x(y)\cdot\One{s(X,y)\leq t}\,\mathrm{d}y\ge\int i_{P,X}(y)\One{s(X,y)\le t}\,\mathrm{d}y=1-\delta(t,X).
\]
Therefore,
\[
\PP{Y_{n+1}\in\Ch^{\mathrm{or}}(X_{n+1})}=\EE{\PPst{s(X,Y)\le t_\alpha}{X}}\geq 1-\EE{\delta(t_\alpha,X)}\ge1-\alpha,
\]
where the last inequality follows from the definition of $t_\alpha$.

We now prove the theorem for the semi-oracle prediction set $\Ch^{\mathrm{sor}}$ in~\eqref{eqn:pred_set_oracle}. Let $\xi(t,X) = \PPst{S > t}{X}$ for $t \in \R$, where $S = s(X,Y)$, and let
\[\tilde{t}_\alpha = \inf\left\{t \in \R : \frac{1}{n+1} \left(\sum_{i=1}^n \delta(t,X_i) + \delta(t,X_{n+1})\right) \leq \alpha\right\},\]
Note that $\tilde{t}_\alpha \leq \hat{t}_\alpha$ holds deterministically, since $\delta(t,X_{n+1}) \leq 1$ deterministically. Now, writing $S_{n+1} = s(X_{n+1},Y_{n+1})$, we have
\begin{multline*}
    \PP{Y_{n+1} \notin \Ch^{\mathrm{sor}}(X_{n+1})} = \PP{S_{n+1} > \hat{t}_\alpha}\\
    = \PP{\bigcup_{k=1}^\infty \left\{S_{n+1} > \hat{t}_\alpha + \frac{1}{k}\right\}} = \lim_{k \rightarrow \infty} \PP{S_{n+1} > \hat{t}_\alpha + \frac{1}{k}}\\
    = \lim_{k \rightarrow \infty} \EE{\PPst{S_{n+1} > \hat{t}_\alpha+\frac{1}{k}}{\hat{t}_\alpha, X_{n+1}}} = \lim_{k \rightarrow \infty} \EE{\xi\left(\hat{t}_\alpha+\frac{1}{k},X_{n+1}\right)},
\end{multline*}
where the last equality holds since $\hat{t}_\alpha$ is independent of $(X_{n+1},Y_{n+1})$ by construction. Next, recall that the density $f_x$ of $P_{Y \mid X=x}$ satisfies \(f_x(y) \geq i_{P,x}(y)\) for all \(x \in \X\) and \(y \in \R\). Therefore, for any $x \in \X$,
\begin{multline*}
    \xi(t,x) = \PPst{S > t}{X = x}  = 1-\PPst{S \leq t}{X=x} = 1 - \int_\R f_x(y) \cdot \One{s(x,y) \leq t} \mathrm{d}y\\
    \leq 1 - \int_\R i_{P,x}(y) \cdot \One{s(x,y) \leq t} \mathrm{d}y = \delta(t,x).
\end{multline*}
Now, since $t \mapsto \delta(t,X)$ is a monotone decreasing function, for any $k = 1,2,\ldots$, we have
\begin{align*}
    &\EE{\xi\left(\hat{t}_\alpha+\frac{1}{k},X_{n+1}\right)} \leq \EE{\delta\left(\hat{t}_\alpha+\frac{1}{k},X_{n+1}\right)} \leq \EE{\delta\left(\tilde{t}_\alpha+\frac{1}{k},X_{n+1}\right)}\\
    &=\frac{1}{n+1}\sum_{i=1}^{n+1} \EE{\delta\left(\tilde{t}_\alpha+\frac{1}{k},X_i\right)} \qquad \text{ since $\tilde{t}_\alpha$ is a symmetric function of $X_1,\ldots,X_{n+1}$}\\
    &= \EE{\frac{1}{n+1}\sum_{i=1}^{n+1} \delta\left(\tilde{t}_\alpha+\frac{1}{k},X_i\right)} \leq \alpha,
\end{align*}
where the last inequality holds by the definition of $\tilde{t}_\alpha$ and the fact that $t \mapsto \delta(t,x)$ is a monotone decreasing function. Putting everything together, we have \(\PP{Y_{n+1} \notin \Ch^{\mathrm{sor}}(X_{n+1})} \leq \alpha\).

\subsection{Proof of Theorem~\ref{thm:finite_measure}}

For the score $s(x,y)=|y|$ and any $t\ge0$, the definition
of $\delta$ gives \(1-\delta(t,x)=\int_{-t}^{t}i_{P,x}(y)\mathrm{d}y\). As $t\to\infty$, this integral increases to $\gamma_P(x)$.
Thus, by the monotone convergence theorem,
\begin{equation*}
    \lim_{t\to\infty}\Ep{X\sim P_X}{\delta(t,X)}
    =1-\Ep{X\sim P_X}{\gamma_P(X)}
    =1-\Gamma_P
    <\alpha.
\end{equation*}
Consequently, there exists a deterministic finite $T\ge0$
such that $\Ep{X\sim P_X}{\delta(T,X)}<\alpha$.
By the definition of the population threshold $t_\alpha$
in~\eqref{eqn:pred_set_oracle_2}, we therefore have
$t_\alpha\le T$.
Note also that $t_\alpha \geq 0$ as long as $\alpha < 1$, since $\delta(t,x)=1$ whenever $t<0$.

It follows that, since $\Ch^{\mathrm{or}}(x)=[-t_\alpha,t_\alpha]$ for every
$x\in\X$, \(\EE{\leb{\Ch^{\mathrm{or}}(X_{n+1})}} = 2t_\alpha \leq 2T <\infty\), which proves the claim.

\subsection{Proof of Lemma~\ref{lem:coverage_dr}}

Throughout the proof, we write $\hat{\psi}_t(\cdot)=\psi_t(\cdot;\hat{\eta})$ for simplicity. Let $B=\sup_{t\in\R}|R_t(\hat{\eta},\eta)|$ and $W=U-L$. By~\eqref{eqn:tail_ineq}, for any $t\in\R$, \(\PP{S>t}\le\EE{\psi_t(O)}\le\EE{\hat{\psi}_t(O)}+B\). Applying Hoeffding's inequality, for any fixed $t\in\R$ and $\epsilon>0$, we have
\[
\PP{\frac1n\sum_{i=1}^n\hat{\psi}_t(O_i)<\EE{\hat{\psi}_t(O)}-\epsilon}
\le\exp\left(-\frac{2n\epsilon^2}{W^2}\right).
\]

If $\alpha+B\ge1$, the desired coverage bound holds trivially. Otherwise, fix any $r\in(\alpha+B,1)$ and define the nonrandom threshold \(t_r^*=\inf\{t\in\R:\PP{S>t}\le r\}\). Since $t\mapsto\PP{S>t}$ is nonincreasing and right-continuous, $t_r^*$ is finite, and $\PP{S>t}>r$ if and only if $t<t_r^*$.

Now, fix any $t<t_r^*$. By the definition of $t_r^*$, \(\EE{\hat{\psi}_t(O)}\ge\PP{S>t}-B>r-B\). Moreover, if $\hat{t}_\alpha<t$, then by the definition of $\hat{t}_\alpha$, $\frac1n\sum_{i=1}^n\hat{\psi}_t(O_i)<\alpha$. Therefore, applying Hoeffding's inequality, we have \(\PP{\hat{t}_\alpha<t} \le\exp\left(-2n(r-\alpha-B)^2/W^2\right)\). Since this holds for any $t < t_r^*$, this implies \(\PP{\hat{t}_\alpha<t_r^*} \le\exp\left(-2n(r-\alpha-B)^2/W^2\right)\). Next, since $\hat{t}_\alpha$ is independent of $S_{n+1}$, we have
\[
\PP{Y_{n+1}\notin\ch(X_{n+1})}
=\EE{S_{n+1}>\hat{t}_\alpha} = \EE{\PPst{S_{n+1}>\hat{t}_\alpha}{\hat{t}_\alpha}} = \EE{Q},
\]
where $Q=\PPst{S_{n+1}>\hat{t}_\alpha}{\hat{t}_\alpha}$. Now, by the definition of $t_r^*$, \(\PP{Q>r} =\PP{\hat{t}_\alpha<t_r^*} \le\exp\left(-2n(r-\alpha-B)^2/W^2\right)\). Consequently, since $Q \in [0,1]$,
\begin{align*}
&\EE{Q}=\int_0^1\PP{Q>r}\,\mathrm{d}r\le\alpha+B+\int_{\alpha+B}^1\PP{Q>r}\,\mathrm{d}r\\
&=\alpha+B+\int_0^{1-\alpha-B}\PP{Q>\alpha+B+u}\,\mathrm{d}u\le\alpha+B+\int_0^\infty
\exp\left(-\frac{2nu^2}{W^2}\right)\mathrm{d}u=\alpha+B+W\sqrt{\frac{\pi}{8n}}.
\end{align*}
This completes the proof.

\subsection{Proof of Proposition~\ref{prop:sharp}}

The upper bound follows directly from the definition of $\delta(t,x)$. To prove the sharpness, we construct a distribution $Q\in\Piv$ that attains the bound.

For each $x$ with $\gamma_P(x)<1$, choose a measurable density $h_{t,x}$ supported on $\{y:s(x,y)>t\}$. Following the construction in the proof of Theorem~\ref{thm:impos}, we construct $Q$ by replacing the point mass $\delta_M$ in the definition of $H_{z,x}$ with the distribution having density $h_{t,x}$. The resulting distribution satisfies $Q_{X,Z,R,RY}=P_{X,Z,R,RY}$ and $Y\indep Z\mid X$.
Moreover, its conditional outcome density given $X=x$ is
\[
f_x^*(y)=i_{P,x}(y)+(1-\gamma_P(x))h_{t,x}(y).
\]
Consequently, for $\gamma_P(x)<1$,
\[
\begin{aligned}
\Ppst{Q}{s(X,Y)>t}{X=x}
&=\int \mathbbm{1}\{s(x,y)>t\}f_x^*(y)\,\mathrm{d}y\\
&=\int \mathbbm{1}\{s(x,y)>t\}i_{P,x}(y)\,\mathrm{d}y
+1-\gamma_P(x)=\delta(t,x).
\end{aligned}
\]
The same equality holds when $\gamma_P(x)=1$, since $f_x^*=i_{P,x}$ in this case. Thus,
\[
\Pp{Q}{s(X,Y)>t}=\Ep{P}{\delta(t,X)},
\]
which establishes the claimed equality.

\subsection{Proof of Lemma~\ref{lem:policy_conditions}}

Consider any sequence $(t_n)_{n\in\N}$ with $\lim_{n\to\infty} t_n=t$. By assumption, for $P_X$-almost every fixed $x\in\X$, \(\One{s(x,y)\le t_n}\longrightarrow\One{s(x,y)\le t}\) almost everywhere. Since 
\[0\le q_{a(x,y),x}(y)\One{s(x,y)\le t_n}\le q_{a(x,y),x}(y) \leq i_{P,x}(y),\]
applying the dominated convergence theorem, we have  $\tilde{\delta}(t_n,x)\longrightarrow\tilde{\delta}(t,x)$
for $P_X$-almost every $x$ (recall that $\int i_{P,x}(y) \mathrm{d}y = \gamma_P(x) < \infty$). Hence, \(\tilde{\delta}(t_n,X) - \tilde{\delta}(t,X) \longrightarrow 0\) almost surely. Since $0\le\tilde{\delta}(t,X)\le1$, another application of the dominated convergence theorem yields \(\EE{\tilde{\delta}(t_n,X) - \tilde{\delta}(t,X)} \longrightarrow 0\), which proves the continuity.

\subsection{Proof of Lemma~\ref{lem:policy_eif}}

By the definition of $m_{z,x}^a(t)$,
\begin{align*}
m_{z,x}^a(t)&=\EE{R\cdot\One{S\le t}\One{a(X,Y)=z}\mid X=x,Z=z}\\
&=\int_{\R}q_{z,x}(y)\One{a(x,y)=z}\One{s(x,y)\le t}\,\mathrm{d}y,
\end{align*}
it follows that
\[
\EE{m_{z,X}^a(t)}
=
\EE{\int_{\R}q_{z,X}(y)\One{a(X,y)=z}\One{s(X,y)\le t}\,\mathrm{d}y}.
\]
Summing over $z\in\Z$ gives
\[
\sum_{z \in \Z}\EE{m_{z,X}^a(t)}
=
\EE{\int_{\R}q_{a(X,y),X}(y)\One{s(X,y)\le t}\,\mathrm{d}y}
=
1-\Psi_{a,t}(P),
\]
where the last equality follows from the definition of $\tilde{\delta}_a$.
Hence,
\begin{align*}
    \EE{\psi_{a,t}(O;\hat{\eta})}-\Psi_{a,t}(P) &=-\sum_{z \in \Z}\EE{\hat{m}_{z,X}^a(t)-m_{z,X}^a(t)}-\sum_{z \in \Z}\EE{\frac{\One{Z=z}}{\hat{\kappa}_{z,X}}\{H_{z,t}^a(O)-\hat{m}_{z,X}^a(t)\}}\\
&=-\sum_{z \in \Z}\EE{\hat{m}_{z,X}^a(t)-m_{z,X}^a(t)} - \sum_{z \in \Z}\EE{\frac{\kappa_{z,X}}{\hat{\kappa}_{z,X}}\{m_{z,X}^a(t)-\hat{m}_{z,X}^a(t)\}}.\\
&=-\sum_{z \in \Z}\EE{\frac{\hat{\kappa}_{z,X}-\kappa_{z,X}}{\hat{\kappa}_{z,X}}\left(\hat{m}_{z,X}^a(t)-m_{z,X}^a(t)\right)},
\end{align*}
which completes the proof. 

\subsection{Proof of Theorem~\ref{thm:policy_coverage}}
By Lemmas~\ref{lem:policy_conditions} and~\ref{lem:coverage_dr}, it suffices to establish that
\[
\sup_{t \in \R} |R_t(\hat{\eta},\eta)| \leq \frac{1}{b}\sum_{z\in\Z}\|\hat{\kappa}_{z,X}-\kappa_{z,X}\|\cdot\sup_{t\in\R}\|\hat{m}_{z,X}^a(t)-m_{z,X}^a(t)\|.
\]
This follows directly from Lemma~\ref{lem:policy_eif}, the condition $\hat{\kappa}_{z,X} > b$, and the Cauchy--Schwarz inequality.

\subsection{Proof of Theorem~\ref{thm:policy_coverage_upper}}

First, by the definition of $\Psi_{a,t}(P)$, for any $t\in\R$, \(\PP{S\le t}=1-\Psi_{a,t}(P)+\Delta_a(t)\), where
\[
\Delta_a(t)=\EE{\int_\R\{f_{Y\mid X}(y)-q_{a(X,y),X}(y)\}\One{s(X,y)\le t}\,\mathrm{d}y}.
\]
Since $0\le q_{a(x,y),x}(y)\le i_{P,x}(y)\le f_x(y)$, we have
$0\le\Delta_a(t)\le1-\Gamma_P+\operatorname{Reg}(a)$.
Moreover, from the proof of Theorem~\ref{thm:policy_coverage}, \(\sup_{t\in\R}\left|\Ep{P}{\psi_{a,t}(O;\hat\eta)}-\Psi_{a,t}(P)\right|\le B_a\). Next, rearranging the EIF in~\eqref{eqn:policy_eif}, we have $\psi_{a,t}(O;\hat\eta)=1-U_t(O)+V_t(O)$,
where
\[
U_t(O)=\sum_{z\ne Z}\hat m_{z,X}^a(t)
+\frac{R\One{a(X,Y)=Z}\One{S\le t}}{\hat\kappa_{Z,X}},
\qquad
V_t(O)=\left(\frac1{\hat\kappa_{Z,X}}-1\right)\hat m_{Z,X}^a(t).
\]
Under Assumption~\ref{asm:policy_upper_shape}, both functions are nondecreasing and right-continuous in $t$, with ranges contained in $[0,|\Z|-1+1/b]$ and $[0,1/b-1]$, respectively. By the expected one-sided Dvoretzky--Kiefer--Wolfowitz bound~\citep[Corollary 2]{reeve2024short}, applied to bounded monotone functions, we have \(\EE{D_n}\le \left(|\Z|+\frac2b-2\right)\sqrt{\frac{\pi}{8n}}\), where
\[
D_n=\left[\sup_{t\in\R}\left\{
\frac1n\sum_{i=1}^n\psi_{a,t}(O_i;\hat\eta)
-\Ep{P}{\psi_{a,t}(O;\hat\eta)}
\right\}\right]_+.
\]
Here, the bound for bounded monotone functions follows by representing each normalized function as the conditional distribution function of an auxiliary random variable and applying Jensen's inequality.

By the definition of $\hat t_\alpha$, every empirically feasible threshold $t$ satisfies \(\Psi_{a,t}(P)\ge \Ep{P}{\psi_{a,t}(O;\hat\eta)}-B_a \ge\alpha-B_a-D_n\). Taking a sequence of feasible thresholds approaching $\hat t_\alpha$ and using the continuity of $t\mapsto\Psi_{a,t}(P)$, we have \(\Psi_{a,\hat t_\alpha}(P)\ge\alpha-B_a-D_n\). The same inequality holds for $\hat t_\alpha=-\infty$ or $+\infty$ by extending $\Psi_{a,t}(P)$ to its corresponding limits, with $\Psi_{a,-\infty}(P)=1$.

Finally, by the independence of the test observation,
\begin{align*}
\PP{Y_{n+1}\in\Ch(X_{n+1})}
&=\EE{1-\Psi_{a,\hat t_\alpha}(P)+\Delta_a(\hat t_\alpha)}\\
&\leq 1-\alpha+B_a+\EE{D_n}+(1-\Gamma_P)+\operatorname{Reg}(a)\\
&\leq 1-\alpha+(1-\Gamma_P)+\operatorname{Reg}(a)+B_a
+\left(|\Z|+\frac2b-2\right)\sqrt{\frac{\pi}{8n}},
\end{align*}
which proves the claim.

\end{document}